\documentclass[%
 aip,
 amsmath,amssymb,floatfix,
 reprint,%
]{revtex4-1}

\usepackage{graphicx}
\usepackage{dcolumn}
\usepackage{bm}

\usepackage[utf8]{inputenc}
\usepackage[T1]{fontenc}
\usepackage{newtxtext,newtxmath}
\usepackage{etoolbox}
\usepackage{xfrac}
\usepackage{float}
\usepackage{color}
\usepackage{umoline}
\makeatletter
\def\@email#1#2{%
 \endgroup
 \patchcmd{\titleblock@produce}
  {\frontmatter@RRAPformat}
  {\frontmatter@RRAPformat{\produce@RRAP{*#1\href{mailto:#2}{#2}}}\frontmatter@RRAPformat}
  {}{}
}%
\makeatother

\begin{document}

\preprint{AIP/123-QED}

\title[Phase Diagram of Flexible Polymers with Quenched
Disordered Charged Particles]{Phase Diagram of Flexible Polymers with Quenched Disordered Charged Monomers}
\author{B.B. Rodrigues}
    \email[Electronic mail:\ ]{bbr@fisica.ufmg.br}
    \affiliation{Laborat\'orio de Simula\c c\~ao, Departamento de F\'isica, ICEx \\  Universidade Federal de Minas Gerais, 31720-901 Belo Horizonte, Minas Gerais, Brazil}
\author{J.C.S. Rocha}
    \email[Electronic mail:\ ]{jcsrocha@ufop.edu.br}
    \affiliation{Departamento de F\'isica, ICEB, Universidade Federal de Ouro Preto, 34000-000 Ouro Preto, Minas Gerais, Brazil}
\author{B.V. Costa}
    \email[Electronic mail:\ ]{bvc@fisica.ufmg.br}
    \affiliation{Laborat\'orio de Simula\c c\~ao, Departamento de F\'isica, ICEx \\ Universidade Federal de Minas Gerais, 31720-901 Belo Horizonte, Minas Gerais, Brazil}
\date{\today}

\begin{abstract}

Recent advances in Generalized Ensemble simulations and microcanonical analysis allowed the investigation of structural transitions in polymer models over a broad range of local bending and torsion strengths. It is reasonable to argue that electrostatic interactions play a significant role in stabilizing and mediating structural transitions in polymers. We propose a bead-spring polymer model with randomly distributed charged monomers interacting via a screened Coulomb potential. By combining the Replica Exchange Wang-Landau (REWL) method with energy-dependent monomer updates, we constructed the hyperphase diagram as a function of temperature ($T$) and charged monomer concentration ($\eta$). The coil-globule and globular-solid transitions are respectively second and first order for the entire concentration range. However, above a concentration threshold of $\eta=80\%$, electrostatic repulsion hinders the formation of solid and liquid globules, and the interplay between enthalpic and entropic interactions leads to the formation of liquid pearl-necklace ad solid helical structures. The probability distribution, $P(E,T)$, indicates that at high $\eta$, the pearl-necklace liquid phase freezes into a stable solid helix-like structure with a free energy barrier higher than the freezing globule transition at low $\eta$.

\end{abstract}

\maketitle

\section{\label{Introduction}Introduction}

The understanding of thermodynamic and mechanical properties of polymers as a response to changes in external parameters (such as temperature, solvent quality, chemical potential, and confinement) or intrinsic interactions is a central problem in several areas, ranging from the production of nanostructured materials~\cite{Angew2007}, development of molecular-engineered pharmaceutical drugs~\cite{Science1992}, protein aggregation~\cite{Takano} and molecular translocation through confined geometries~\cite{KUMAR2018216}. Minimalist models with local interactions (bonding, bending, and torsion potentials) have shed light on mapping phase transitions in macromolecules~\cite{MichaelPRL}. Of particular interest to understand collective processes, models with non-local interactions, such as hydrophobic-polar~\cite{THIRUMALAI2003146}, polyampholytes~\cite{Polyampholyte2004}, and polyelectrolytes~\cite{Polyelectrolyte2005}, have unraveled the formation of secondary and tertiary structures~\cite{Bloomfield} with underlying mechanisms that are still under debate in Literature~\cite{Dobrynin2021}.

Advances in experimental techniques allowed one to manipulate systems as small as one macromolecule, although the access to observables related to transition pathways is still unattainable~\cite{lipfert2014understanding,guerois2002predicting}. Computationally, the identification of stable conformations as well as classifying structural transitions employing Monte Carlo (MC) simulations became more affordable with the advent of generalized ensemble methods, such as multicanonical (MUCA)~\cite{Berg2} and Wang-Landau (WL)~\cite{LandauPRL}. Based on the parallel tempering algorithm\cite{Swendsen}, multi-thread replica exchange methods~\cite{REWL} increased the scalability by dividing the energy range into small overlapping intervals~\cite{farris2021replica}. We can push the boundaries in sampling by considering energy-dependent local~\cite{Schnabel} and cluster\cite{Kampmann} MC moves. The direct sampling of the density of states (DOS), $g(E)$, also opens the possibility to study phase transitions in finite systems at the microcanonical ensemble~\cite{ispolatov2001first,Binder2005}, by identifying non-analyticies in $g(E)$ and its derivatives~\cite{MichaelPRL2018}, finding zeros from the partition function~\cite{Julio-1}, or the energy probability distribution~\cite{costa2017energy,rodrigues2022pushing}.

Coarse-grained models for polymers have been extensively applied as benchmarks to test and improve generalized ensemble algorithms~\cite{Janke}. Such simulations contributed to accurately locating and classifying phase transitions in flexible, semiflexible, and stiff~\cite{MichaelPRL,SeatonPRL} diluted single chains, along with refining predictions for the coil-globule transition temperature in the thermodynamic limit, the so-called $\Theta-$point, $T_{\Theta}(N\rightarrow \infty)$\cite{PolymerPhysics,seatonPRE}. Besides that, they also helped to understand the role of bonded~\cite{Achille} and non-bonded~\cite{Michael2013} interactions on the order of both collapse (random coil to a liquid globular) and freezing (liquid to solid globule) transitions in flexible polymers. Such studies also settled the conditions at which polymers fold into solid toroidal and helix-like structures due to bending and torsional restraints~\cite{MichaelPRL}. In these cases, the suppression of entropic degrees of freedom hinders the occurrence of first-order freezing transitions~\cite{aierken2020comparison}. In this finite-system framework, transitions driven by the interplay between entropic and pairwise energetic contributions (such as electrostatic interactions) are still unclear~\cite{jacs2018}. It is not far-fetched to infer that another step towards modeling folding transitions in macromolecules is the inclusion of non-bonded repulsive interactions. It is not far-fetched to infer that a step towards understanding folding transitions in macromolecules will account for the effect of non-bonded repulsive interactions.

In this work we consider a minimalist bead-spring polymer model with short-ranged repulsive monomers randomly placed along the chain for a wide range of concentrations. By accurately sampling $g(E)$ with REWL and computing the quenched-disordered canonical fluctuations of thermal and structural observables, we obtain the $(T,\eta)$ hyperphase diagram. Although the tendency of the collapse and freezing transition temperatures to merge at high strength interactions shares similarities with bending stiffness models~\cite{SeatonPRL}, microcanonical analysis indicates that, in our model, the collapse transition remains second-order. We also compute the probability distribution, $P(E,T)$, and demonstrate that the freezing transition remains first-order at high $\eta$, with the liquid pearl-necklace structures folding into helix-like solid configurations, separated by a high free energy barrier.

The paper is organized as follows. Section~\ref{Model} describes the proposed polymer model. Both \mbox{REWL} and energy-dependent MC moves are briefly addressed in section~\ref{Simulation_Details}, followed by a description of how we calculate thermal and quenched disordered averages. Section~\ref{results} presents our results for the hyperphase diagram and the classification of the coil-globule, liquid-solid, and solid-solid transitions. Finally, in section~\ref{Conclusions} we summarize the conclusions and suggest further investigations not addressed in this study.

\section{\label{Model}Model}

We chose a minimalist approach for our computational experiment to model the polymer. As usual, we treat the monomers as spherical beads, interacting through the extensively used Lennard-Jones (6-12)
\begin{equation}
    V_{\text{LJ}}\left(r_{ij}\right) = 4\epsilon\displaystyle\left[\left(\frac{\sigma}{r_{ij}}\right)^{12} - \left(\frac{\sigma}{r_{ij}}\right)^{6}\right] - V_{\text{LJ}}\left(r_{1}\right)   ~~~,
\label{Lennard-Jones}
\end{equation}
\noindent
 and the anharmonic FENE (Finitely Extensible Nonlinear Elastic) potentials:
\begin{equation}
    V_{\text{F}}\left(r_{i,i+1}\right) =- \frac{1}{2}k_0R^2\ln \left[1 - \left(\frac{r_{i,i+1} - r_{0}}{R}\right)^2\right]   ~~~,
    \label{FENE}
\end{equation}
\noindent
where $r_{ij}$ is the distance between monomers at positions $i$ and $j$,  $\sigma=2^{-1/6}r_0$ is the effective radius of the monomer in units of $r_0$, which is the minimum of the \mbox{LJ}  potential and sets the reduced length unit. A cutoff at $r_{1} = 2.5\sigma$ is introduced, and $V_{\text{LJ}}$ is shifted by $V_{\text{LJ}}\left(r_{1}\right)\approx -10^{-6}$.  For $r_{ij}>r_{1}$, $V_{\text{LJ}}\left(r_{ij}\right)=0$. The potential depth $\epsilon$ is set to unity and energy is given in dimensionless units $E/\epsilon$. In $V_{\text{F}}$ we set the stiffness constant $k_0=40$ and  the finite extensibility $R=3/7$, as commonly used for flexible polymers~\cite{seatonPRE}.

\begin{figure}[htp]
\centering
\includegraphics[width=0.45\textwidth]{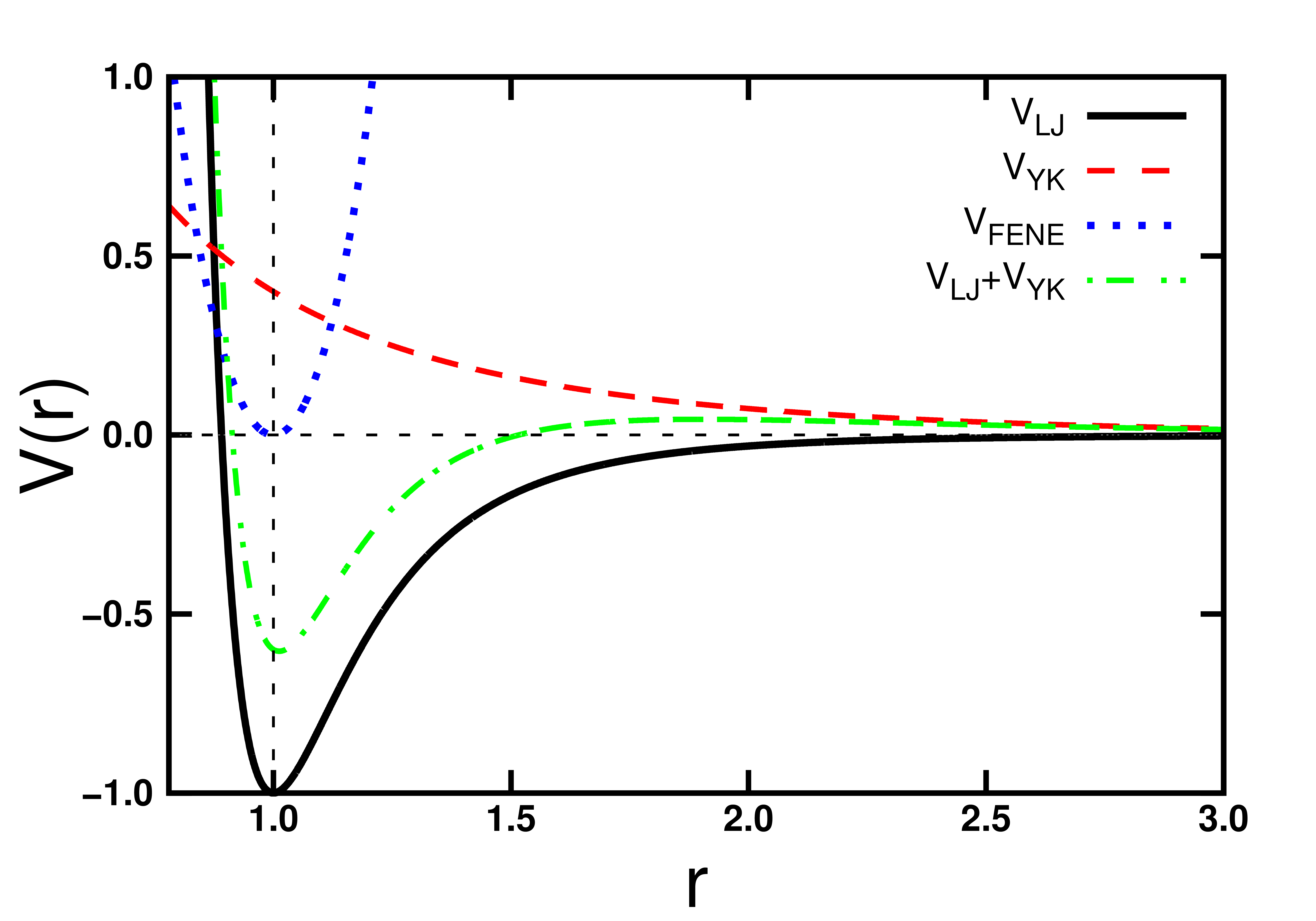}
\caption{(Color online) Interacting potentials of the coarse-grained model: Lennard-Jones (black solid curve), Yukawa (red dashed curve), FENE (dotted blue curve) and LJ+YK combined (green dot-dashed curve).}
\label{Fig_potential}
\end{figure}

Particles with an excess charge randomly irreversibly attach to the polymer chain. We assume that these monomers have one unit charge, and the implicit solvent screens the Coulomb repulsion, $V_{\text{YK}}$ 
\begin{equation}
V_{\text{YK}}\left(r_{ij}\right)=\frac{\alpha^2}{r_{ij}}e^{- r_{ij}/\xi} - V_{\text{YK}}\left(r_{2}\right),
\label{yk}
\end{equation}
\noindent    
where $\alpha^2$  is the Bjerrum length parameter and $\xi$ is the Yukawa screening length. We set $\alpha^2=0.4$ and $\xi=1.0$, to match the physiological condition where electrostatic interactions are strongly screened~\cite{dobrynin2008theory}. A cutoff, $r_2=5\sigma$, is introduced in order to account for the screened Coulomb repulsion.  The interacting potentials are shown in Fig. \ref{Fig_potential}.

The total energy of the polymer with $N$ monomers is given by
 \begin{equation}
    \mathcal H =   \sum_{i=1}^{N-1}V_{\text{F}} \left(r_{i,i+1}\right)  + \sum_{i=1}^{N-1}\sum_{j=i+1}^{N}\left[ V_{\text{LJ}}\left(r_{ij}\right)+V_{\text{YK}}\left(r_{ij}\right)\right].
\label{hamiltonian}
\end{equation}
\noindent
In this work we performed simulations on chains with $N=20$, $50$ and $70$ monomers. The smaller chain folds into a liquid globular phase for all $\eta$ values, and the intermediate size ($N=50$ monomers) chain produced results similar to $N=70$, which are the ones reported here.

\section{Simulation Details and Microcanonical Analysis}
\label{Simulation_Details}
\noindent

\subsection{Simulation Details}

We conducted independent simulations of a single chain with a fixed number of constituents, $N$, and  charged monomers concentration ranging from $\eta=0$ to $100\%$, incremented by $10\%$, in an infinitely diluted volume. From a statistical point of view, the density of states, $g(E)$, is the \emph{sine qua non} quantity, in the sense that it carries all thermodynamic information of the system. The Wang-Landau (WL) and the multicanonical (MUCA) methods belong to a class of algorithms with specific schemes designed to estimate $g(E)$. In those algorithms the acceptance probability of a trial state is inversely proportional to the DOS, hence,  by following the Metropolis recipe, it can be written as $\text{W}_{i\to f} =\text{min}\left\{\frac{g\left(E_i\right)}{g\left(E_f\right)},1\right\}$. We implemented the following trial moves for the monomers in the chain: energy-dependent random displacement, crank-shaft, and pivotation~\cite{Stefan2009}. Here one Monte Carlo sweep (MCS) represents a batch of MC steps that comprise: $N$ single monomer random displacements and one of each rotational moves.

In the original WL method, a guess of the initial entropy is set as $\ln{g(E)} = 0$. Hereinafter $g(E)$ is updated by a multiplicative factor $f$, i.e $\ln{g(E)} \leftarrow \ln{g(E)} + \ln{(f)}$, every time a state with energy between $E$ and $E +\delta E$ is sampled, in this work we considered $\delta E = 0.05$. Simultaneously, one should count how many states in that energy bin were sampled, i.e  $h(E) \leftarrow h(E) + 1$. By construction, when all energies bins are equally sampled, i.e the histogram is flat, $h(E)= c$, the estimated DOS approaches to the exact value, up to a multiplicative constant, with precision equal to $\sqrt{f}$.  At first, one can choose a large value for $f = f_0$ to faster sample the entire configuration space (we chose the original prescription $\ln{(f_0)} = 1$). The histogram flatness is tested once and again after $MCS = 10^5 n_{\mathrm{bin}}$. If the histogram is not flat we  continue the process with $10\%$ of the $MCS$. Else, $f$ is decreased, in order to improve the precision, the histogram is reset, $h(E) = 0$, and the scheme is repeated. The histogram is considered flat when the ratio of its lowest value by the mean value is greater than a threshold $p$, in this work we set $p=70\%$. Any function can be used to decrease $f$, at first we also used the original instruction, i.e. $\ln{(f_{i+1})} = \ln{(f_i)}/2$. The simulation runs until the desired precision is reached, here we cease the process when $\ln{(f)} = \varepsilon = 10^{-8}$. We considered the energy ranging from $E_{\text{min}}$ to $E_{\text{max}}/N=1.0$. Since the ground energy is not known, $E_{\text{min}}$ is calculated iteratively, which leads to $n_{\mathrm{bin}}\propto 10^4$ energy bins.

Since the precision is proportional to $\sqrt{f}$ a cumulative error on the estimate of $g(E)$ is observed if $f$ remains unchanged for a long simulation time. In order to overcome this issue, we implemented the $1/t$ modification factor rule~\cite{belardinelli2007fast}, where $t=j/n_{\mathrm{bin}}$ is the normalized time of simulation, and $j$ is the number of trial moves performed. In this procedure the regular WL scheme is unaltered while $\ln{(f)} > 1/t$. Thereafter, after every $n_{\mathrm{bin}}$ trial move, i.e $t \leftarrow t + 1$, the multiplicative factor is updated as $\ln{(f_{i+1})} = \ln{(f_i)}/t$. This continuously slight modification on $f$ allows a fine-tuning of $g(E)$ with precision proportional to $1/\sqrt{t}$.

The standard WL method is very time-consuming, and it is prohibitive for many problems. However, a parallelization procedure can turn an intractable problem into a feasible one. The idea is to divide the energy range into several smaller pieces, called windows, and one or more WL schemes, called walkers, are performed in parallel at each window. At the end of the process, the pieces of the entropies are combined to form the entire DOS\cite{MichaelBook}. Besides that, an attempt to exchange configurations of walkers between adjacent windows is proposed after $N_{WL}\times n^i_{bin}$ MC moves. An exchange between conformations $X_i$ and $X_j$, respectively located at neighboring windows $i$ and $j$, is proposed with the probability
 \begin{equation}
     \text{P}_{\text{acc}} =\text{min}\left\{\frac{g_i\left(E\left[X_i\right]\right)}{g_i\left(E\left[X_j\right]\right)}\frac{g_j\left(E\left[X_j\right]\right)}{g_j\left(E\left[X_i\right]\right)},1\right\}.
 \end{equation} 
 
This scheme is called Replica Exchange Wang-Landau (REWL) method~\cite{REWL}. Due to the effectiveness of the replica exchange to sample the configuration space, this procedure is as crucial as dividing the windows to improve the simulation time. The replica exchange acceptance ratio is tied to the overlap between the windows. In this work, we considered an overlap of $80\%$, which led to an acceptance higher than $50\%$. We set $4$ independent walkers in each window with the number of windows ranging between $8-12$, depending on the system size. 
\subsection{Canonical and Microcanonical Analysis}
 We sampled thermal and structural observables after $g\left(E\right)$ converges, in a production run where each energy bin is visited at least $10^7$ times. The relevant chain size quantities are the radius of gyration
\begin{equation}
    R_{\mathrm{gyr}}^2 = \frac{1}{N}\sum\limits_{i=1}^{N}\left(\vec{r}_i - \vec{r}_{\mathrm{CM}}\right)^2,
    \label{rgyr}
\end{equation}
\noindent
where $ \vec{r}_{\mathrm{CM}}$ are the coordinates of the center of mass of the polymer, the squared end-to-end distance
\begin{equation}
    R_{\mathrm{ee}}^2 = \left(\vec{r}_N - \vec{r}_{1}\right)^2,
        \label{ree}
\end{equation}
\noindent
and the contour length
\begin{equation}
R_{\mathrm{c}} = \sum\limits_{i=1}^{N-1}\left|\vec{r}_{i+1} - \vec{r}_{i}\right|.
    \label{rc}
\end{equation}
\noindent
 For a chain with $N$ monomers and concentration $\eta (\%)$ we randomly distribute $N_{\ell} = \eta\times N$ charged monomers along the chain. Let us define a label $x$ as follow: If the $i^{th}$ monomer is charged $x_i=1$ otherwise $x_i=0$. Thus, the calculation of the \mbox{YK} interaction in Eq. \ref{hamiltonian} is evaluated between pairs of monomers labeled with $x_i=1$.  Hence,  the probability of a given configuration $X=\{ x_i \}$ to be generated is $P\left[X\right] = \sfrac{1}{\binom{N}{N_{\ell}}}$. We set $M$ independent initial configurations for every value of $\eta$. Understandably, each initial configuration leads to a distinct DOS, $g\left(E\left[X\right]\right)$, and a set of microcanonical observables, $A\left(E\left[X\right]\right)$. The microcanonical entropy is 
\begin{equation}
    S\left(E\left[X\right]\right) \equiv k_B\ln\left[g\left(E\left[X\right]\right)\right]   ~~~.
\label{Entropy}
\end{equation}
\noindent

We represent the canonical average with brackets, $\left<\dots\right>$, to differentiate from the average over the quenched disorder, denoted by the overline, $\overline{\left<\dots\right>}$. Due to the pairwise nature of the interaction between the quenched disordered monomers, each individual configuration $X=\{ x_i \}$ is translationally invariant, thus the averages over the disorder converge with few independent simulations. The partition function for each DOS is
\begin{equation}
    Z \left[X\right]_{\mathrm{T}} = \sum\limits_{E\left[X\right]}g\left(E\left[X\right]\right)e^{-\beta E\left[X\right] }.
\end{equation}
\noindent
The average of an observable $A\left[X\right]$ at the canonical ensemble for an independent initial configuration is defined as
\begin{equation}
    \left<A\left[X\right]\right>_{\mathrm{T}} = \frac{\sum\limits_{E\left[X\right]}A\left(E\left[X\right]\right)g\left(E\left[X\right]\right) e^{-\beta E\left[X\right] }}{Z \left[X\right]_{\mathrm{T}}}.
\end{equation}
\noindent
This quantity is then averaged over all $M$ independent realizations of the disorder, resulting in
\begin{equation}
    \overline{\left<A\left[X\right]\right>}_{\mathrm{T}} = \sum\limits_{M}P\left[X\right]\left<A\left[X\right]\right>_{\mathrm{T}}.
\label{eq:average}
\end{equation}
\noindent
The free energy, $F\left(T\right)$, is calculated as~\cite{Grinstein}
\begin{eqnarray}
F\left(T\right) & = &  -\beta^{-1} \overline{\ln Z\left[X\right]}_{\mathrm{T}},\nonumber\\
                      & = &  -\beta^{-1}\sum\limits_{X}P\left[X\right]\ln Z\left[X\right]_{\mathrm{T}}.
\label{eq:free}
\end{eqnarray}
\noindent
 We obtained  the error bars from the average over different realizations of the disorder. For the sake of clarity, we set the brackets notation, $\langle\dots\rangle$, as representing both thermal and quenched averages.  In the canonical ensemble, fluctuations of energetic and structural quantities are a signal for locating phase transitions. The fluctuation of an observable $A\left[X\right]$ is the temperature derivative in the canonical ensemble: 
\begin{equation}
    \frac{\mathrm{d} }{\mathrm{d}T}\left<A\left[X\right]\right>_{\mathrm{T}} = 
     \beta^{-2}\left[ \left< A\left[X\right] E\left[X\right] \right>_{\mathrm{T}} -  \left<A\left[X\right]\right>_{\mathrm{T}} \left<E\left[X\right]\right>_{\mathrm{T}}\right],
\label{drgyr}
\end{equation}
where the observables are the ones defined in Eqs. \ref{rgyr}$-$\ref{rc}. The specic heat is the fluctuation of the energy as the observable in Eq. \ref{drgyr}.

The microcanonical inverse temperature is defined as
\begin{equation}
    \beta_{\mathrm{mc}} \equiv \beta\left(E\right) := \frac{\mathrm{d}S\left(E\right)}{\mathrm{d}E}   ~~~.
\label{eq:temp_micro}
\end{equation}
\noindent
A subscript is included to differentiate from the canonical temperature, defined as
\begin{equation}
    \beta_{\mathrm{can}} \equiv \beta\left(T\right) := \frac{1}{k_BT}.
\label{eq:temp_can}
\end{equation}
\noindent
With both equations \ref{eq:free} and \ref{eq:temp_can}, the average of any quantity, $A\left(E\right)$, in the canonical ensemble can be calculated as
\begin{eqnarray}
    \left< A(T)\right> & = & \displaystyle\sum\limits_{E}A(E)e^{ \ln g\left(E\right)-\beta_{\mathrm{can}} \left[E - F\left(T\right)\right] }\nonumber\\
                                    & = & \displaystyle\sum\limits_{E}A(E)P\left(E,T\right)
\label{eq:average_can}
\end{eqnarray}
\noindent
    $P\left(E,T\right)$ is the canonical probability distribution. For a given temperature, the energy that maximizes the probability is defined as the equilibrium energy. Important quantities are obtained as derivatives of the logarithm of $\ln P\left(E,T\right)$
\begin{equation}
    -\ln P\left(E,T\right) = \ln g\left(E\right)-\beta_{\mathrm{can}} \left[E - F\left(T\right)\right]   ~~~.
\end{equation}
\noindent
    with the first derivative with respect to the energy being
\begin{equation}
    \frac{\mathrm{d}}{\mathrm{d}E}\left[-\ln P\left(E,T\right)\right] = \beta_{\mathrm{can}}\left(T\right) -\beta_{\mathrm{mc}}\left(E\right),
\label{eq:condition1}
\end{equation}
\noindent
and the corresponding second derivative
\begin{equation}
    \frac{\mathrm{d}^2}{\mathrm{d}E^2}\left[-\ln P\left(E,T\right)\right] = -\gamma_{\mathrm{mc}}\left(E\right),
\label{eq:condition2}
\end{equation}
\noindent
    where $\gamma_{\mathrm{mc}}$ is the fugacity. In a first-order transition, the probability distribution has three critical points where its first derivative is zero (two maxima an a minimum in between)
\begin{equation}
    \frac{\mathrm{d}}{\mathrm{d}E}\left[-\ln P\left(E,T\right)\right] = 0.
\label{eq:condition3}
\end{equation}
\noindent
The solution of the equation above is
\begin{equation}
    \beta^{tr}_{\mathrm{mc}}\left(E_1\right) = \beta^{tr}_{\mathrm{mc}}\left(E_2\right)  =  \beta^{tr}_{\mathrm{mc}}\left(E_3\right) = \beta^{tr}_{\mathrm{can}}\left(T^{tr}\right)   ~~~,
\label{eq:beta_cond}
\end{equation}
\noindent
with the following conditions for the second derivatives
\begin{eqnarray}
    \gamma^{tr}_{\mathrm{mc}}\left(E_1\right) & < & 0, \quad \text{maximum}\nonumber\\
    \gamma^{tr}_{\mathrm{mc}}\left(E_2\right) & > & 0, \quad \text{minimum}\\
    \gamma^{tr}_{\mathrm{mc}}\left(E_3\right) & < & 0,\quad \text{maximum}.\nonumber
\label{eq:gamma_cond}
\end{eqnarray}
\noindent
These solutions require a positive value for the fugacity at some point between the two negative values. This point also represents the energy, $E$, at which the probability distribution has its lowest value between the two phases $A$ and $B$. In a canonical simulation, sampling around this region is strongly suppressed, due to the difference between the minimum and the peak probabilities posing as a thermodynamic barrier. We observe that this condition is the same as the first order microcanonical condition defined in Ref.~\cite{MichaelBook}.

Continuous transitions are connected to peaks on the second order derivative of the free energy, hence connected to higher order derivatives of the probability distribution. Indeed, an interpretation of the microcanonical analysis can be made as follows. A continuous transition lacks a phase coexistence, the probability function does not display a bimodal distribution. Likewise, no discontinuity is observed in the canonical energy as no latent heat is involved.  Moreover, for finite systems, it is reasonable to argue that a range of temperature characterizes the transition. Therefore,a wide probability distribution for temperatures close to transition $T^{tr}$  is expected. On the other hand, for $T_A < T^{tr}$, i.e. for the lower energy phase $A$, and  for $T_B > T^{tr}$, i.e. for the higher energy phase $B$, the probability distribution is supposed to be narrower, leading to a higher peak in comparison with the transition temperature (considering the total probability is $\int_EP\left(E,T\right)=1$). The maximum of $P\left(E,T\right)$ as a function of the temperature is defined as
\begin{equation}
    F^{\mathrm{MAX}}\left(E,T\right)  = \mathop{\boldsymbol{\max}} \left[-\ln P\left(E,T\right)\right].
\label{eq:fmax}
\end{equation}
\noindent
This condition was already defined as $\beta^{tr}_{\mathrm{mc}}\left(E\right) = \beta^{tr}_{\mathrm{can}}\left(T\right)$ and $\gamma^{tr}_{\mathrm{mc}} <0$. However, in this case we collect only one point for each temperature. Introducing this condition at equation \ref{eq:fmax} leads to
 \begin{equation}
    F^{\mathrm{MAX}}\left(E,T\right)  = \ln g\left(E\right)-\frac{\mathrm{d}\ln g\left(E\right)}{\mathrm{d}E} \left[E - F\left(T\right)\right]
\end{equation}
\noindent
and the location of the maximum for a given temperature is obtained by the partial derivative with respect to the energy
\begin{eqnarray}
    \frac{\partial F^{\mathrm{MAX}}\left(E,T\right)}{\partial E} & = & -\frac{\mathrm{d}^2\ln g\left(E\right)}{\mathrm{d}E^2} \left[E - F\left(T\right)\right] = 0\nonumber\\
    & = &- \gamma\left(E\right)\left[E - F\left(T\right)\right] = 0.
\label{eq:fmax_linha}
\end{eqnarray}
\noindent
The conditions at which the derivative is zero are: $i)$  $\gamma\left(E\right)=0$, which does not lead to a maximum at the probability distribution; and $ii)$ $E - F\left(T\right)= 0$, where $E$ is the microcanonical energy. A derivative of equation \ref{eq:fmax_linha}, at the condition that maximizes the probability leads to
\begin{equation}
\frac{\partial^2 F^{MAX}(E,T)}{\partial E^2}   = - \gamma\left(E\right).
\label{eq:fmax_linha2}
\end{equation}
\noindent
So, for the initial assumption to hold (the peak distribution has a minimum value at $T=T^{tr}$ in comparison with adjacent temperatures), its second derivative must be positive, $\gamma\left(E\right) < 0$, which coincides with the microcanonical condition for the fugacity.

\section{Results and discussion}
\label{results}
\noindent

In this section, we analyze the influence of $\eta$ as an external parameter on the phase transitions of diluted polymer chains of size $N=70$ monomers. We obtain the hyperphase diagram in $(T,\eta)$ from the fluctuations in thermal and structural quantities in the canonical ensemble. We then perform microcanonical analysis to confirm the canonical results and classify both coil-globule and liquid-solid transitions, and investigate the influence of $\eta$ on the free energy barrier at the freezing transition.

\subsection{Structural and energetic observables in the canonical ensemble}
\label{Therma_Fluctuaions}
\noindent

Here we discuss the changes in polymer characteristic sizes followed by the fluctuations in thermal and structural quantities for chains with $N=70$ monomers for the range of charged monomers concentration studied. These fluctuations are a proxy for locating phase transition temperatures.

\begin{figure}[htp]
\begin{tabular}{c}
    \includegraphics[width=0.45\textwidth,keepaspectratio=true]{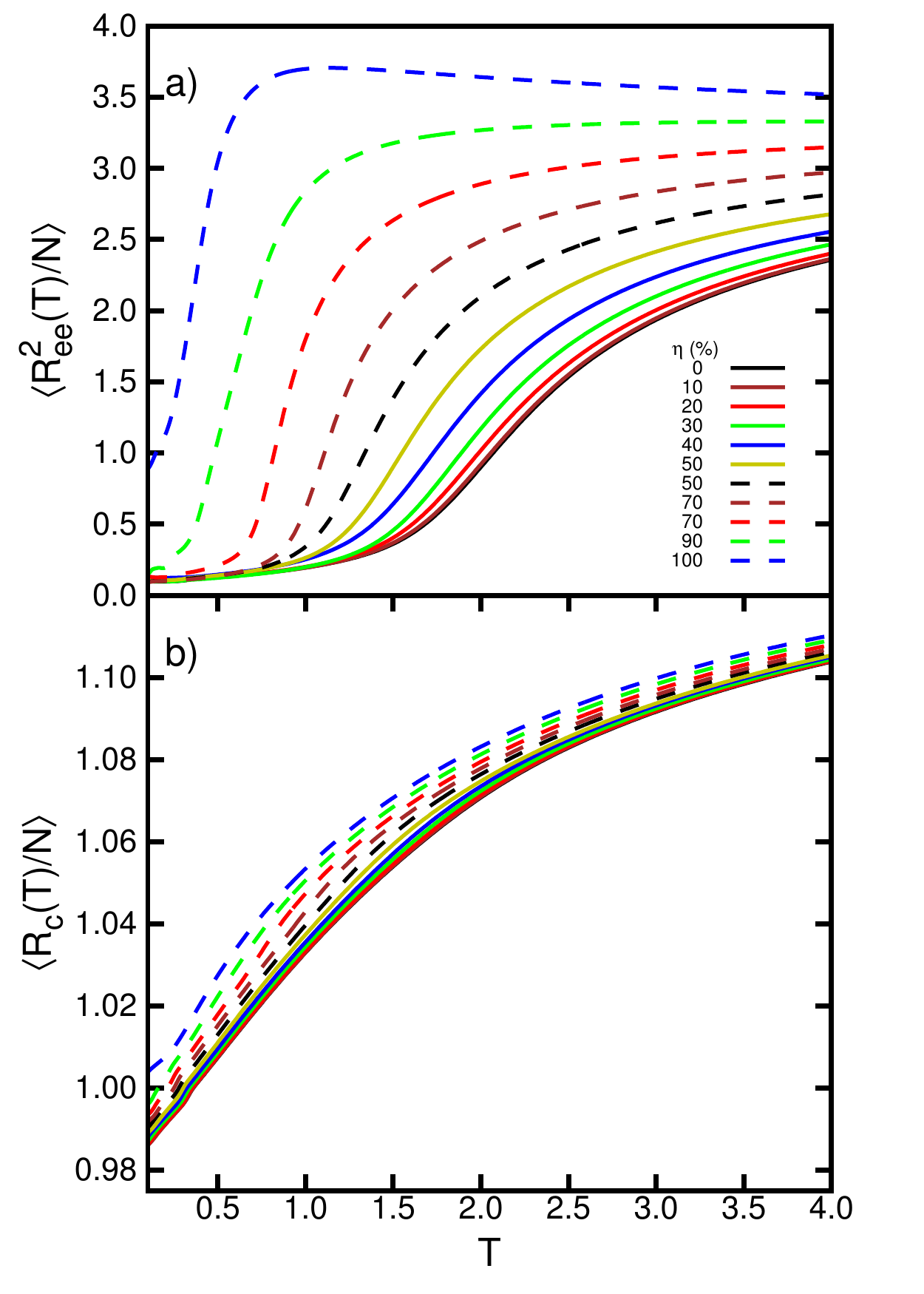}
\end{tabular}
\caption{(Color online) Squared end-to-end distance $(a)$ and contour length $(b)$ per monomer as a function of temperature for $N=70$ and $\eta$.}
\label{Fig3a_structural}
\end{figure}

In Fig.~\ref{Fig3a_structural}~$(a)$ we show the squared end-to-end distance, $\langle R^2_{ee}\left(T\right)\rangle$, and in Fig~\ref{Fig3a_structural}~$(b)$ we present the contour length, $\langle R_{c}\left(T\right)/N\rangle$, in function of temperature, $T$. The end-to-end distance is very small at low $T$, indicating a collapsed phase. Both quantities increase with the temperature until reaching a plateau at the extended phase. Our results show that the end-to-end distance in the extended phase increases with $\eta$, while the temperature of the collapsed transition lowers towards the freezing transition value. We find similar behavior in the squared radius of gyration, $\langle R^2_{gyr}\left(T\right)\rangle$, a plot is not present here to avoid redundancy. The contour length (per monomer-monomer bond unity) increases monotonically with $T$ without a sharp transition between the extended and collapsed phases, as the FENE potential binds the bond length. Noticeably, at low $T$ and high $\eta$, both end-to-end and contour length sizes have a larger characteristic size than the lower concentration curves. This deviation signals that the collapsed conformation is less compact than the expected liquid globule. Thus, the resulting phase might result from the interplay between surface and entropic energies, which will become apparent in the microcanonical analysis.

\begin{figure}[htp]
    \includegraphics[width=0.45\textwidth,keepaspectratio=true]{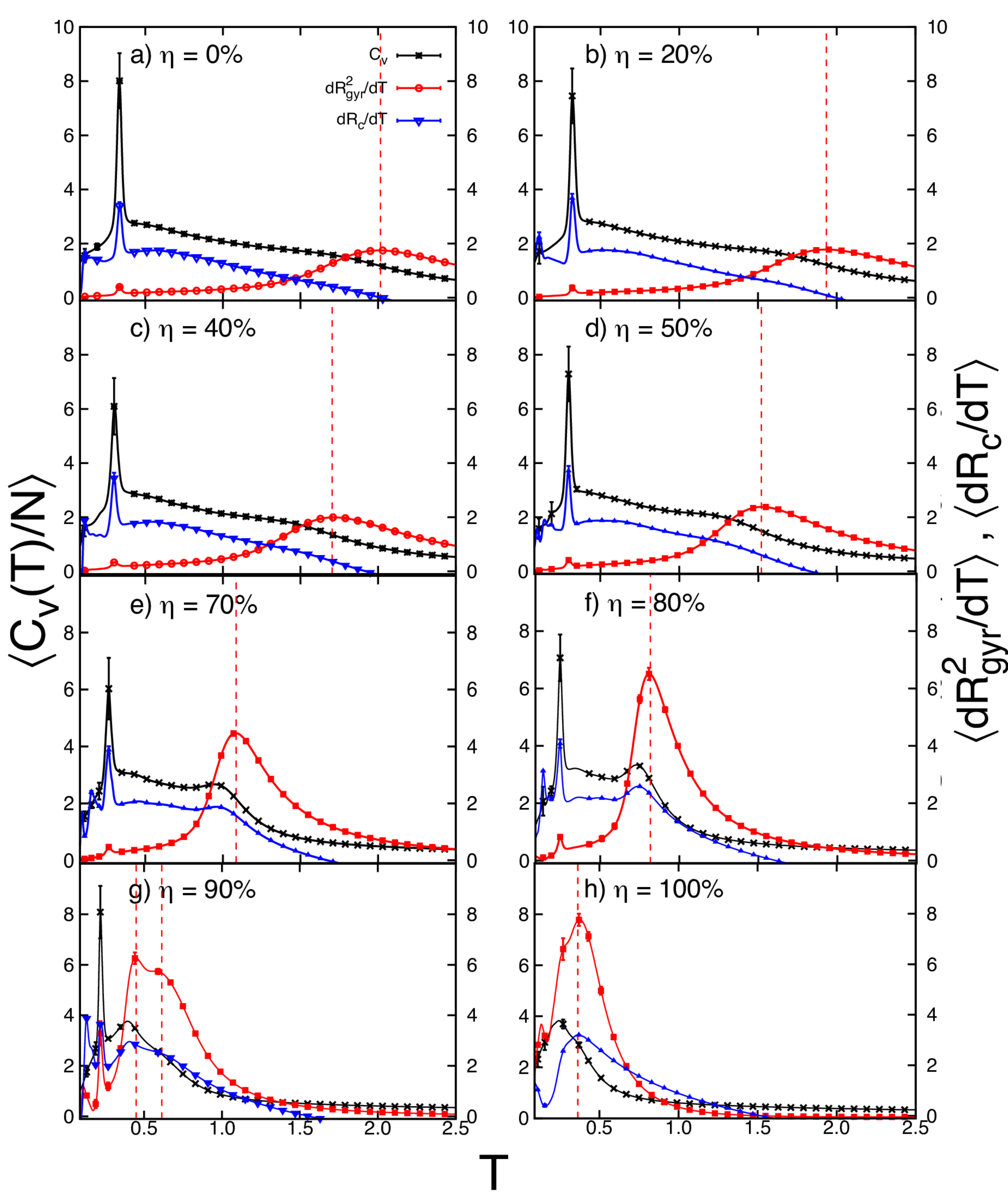}
    \caption{(Color online) Specific heat (black-squared) and thermal fluctuations of radius of gyration (red-circular) contour length (blue-triangular) as a function of temperature for $N=70$ monomers for selected values of $\eta$. Error bars are smaller than the symbols, when not shown.}
    \label{Fig3_critical}
\end{figure}

In Fig.~\ref{Fig3_critical}, we present the specific heat (black-squared curve), the fluctuations of the radius of gyration (red-circular curve), and of the the contour length (blue-triangular curve) for selected values of $\eta$ in panels $a-h$. Shoulders (or less pronounced peaks) at high $T$ signify a transition from an extended to a collapsed phase, while sharp peaks at lower $T$ indicate a freezing transition from the collapsed to a solid-like amorphous structure. In the literature, both transitions merge if either angle-restricting~\cite{seatonPRE} or bonded~\cite{koci2015confinement} interactions are strong enough.

The increase in charged monomer density leads to the approximation of the collapsed transition temperature towards the freezing value. At $\eta=0\%$, both transitions are well separated, while for $\eta=90\%$, the collapse transition occurs closer to the freezing temperature, and look indistinguishable in Fig.~\ref{Fig3_critical}. By introducing an energetic penalty due to repulsion but still allowing access to a minimum in the LJ interaction, the model becomes an aggregation problem, as recently reported by Taylor et. al~\cite{Binder2016} and Truguilho and Rizzi~\cite{zierenberg2016dilute}. We will see from the microcanonical analysis that both transitions are unambiguously identified and the freezing transition is first-order independently of the  concentration.

For $\eta=90\%$, we can identify two nearby peaks in both energetic and structural fluctuations, see Fig.~\ref{Fig3_critical}-$(g)$. The double peaks are actually the result of a two-step collapse, as the chain partially folds into a to pearl-necklace structure.  For $\eta=100\%$, the collapse transition still occurs, however the globular-like structure is no longer stable. The energetic penalty at the collapsed phase is mainly due to attractive reorganizations, therefore, the repulsion between charged monomers have little influence. 

The liquid-solid transition sill occurs for $\eta \geq 90\%$ and is identified by a small peak in $\langle \mathrm{d} R_{\mathrm{c}}/\mathrm{d}T\rangle$ at low $T$. The unfolding of liquid-like pearl-necklace structures into a solid helix-like conformation is driven by the interplay between surface interactions and all competitive enthalpic interactions, as the pairs of repulsive monomers interacting via combination of attractive LJ at short distances and a repulsive YK at intermediate distances. There are also smaller signals at very low $T$, arising from further accommodation of surface contacts of the solid structure and minimization of entangled parts of the chain.

\subsection{Hyperphase diagram}
\label{pd}

\begin{figure}[htp]
    \includegraphics[width=0.4\textwidth,keepaspectratio=true]{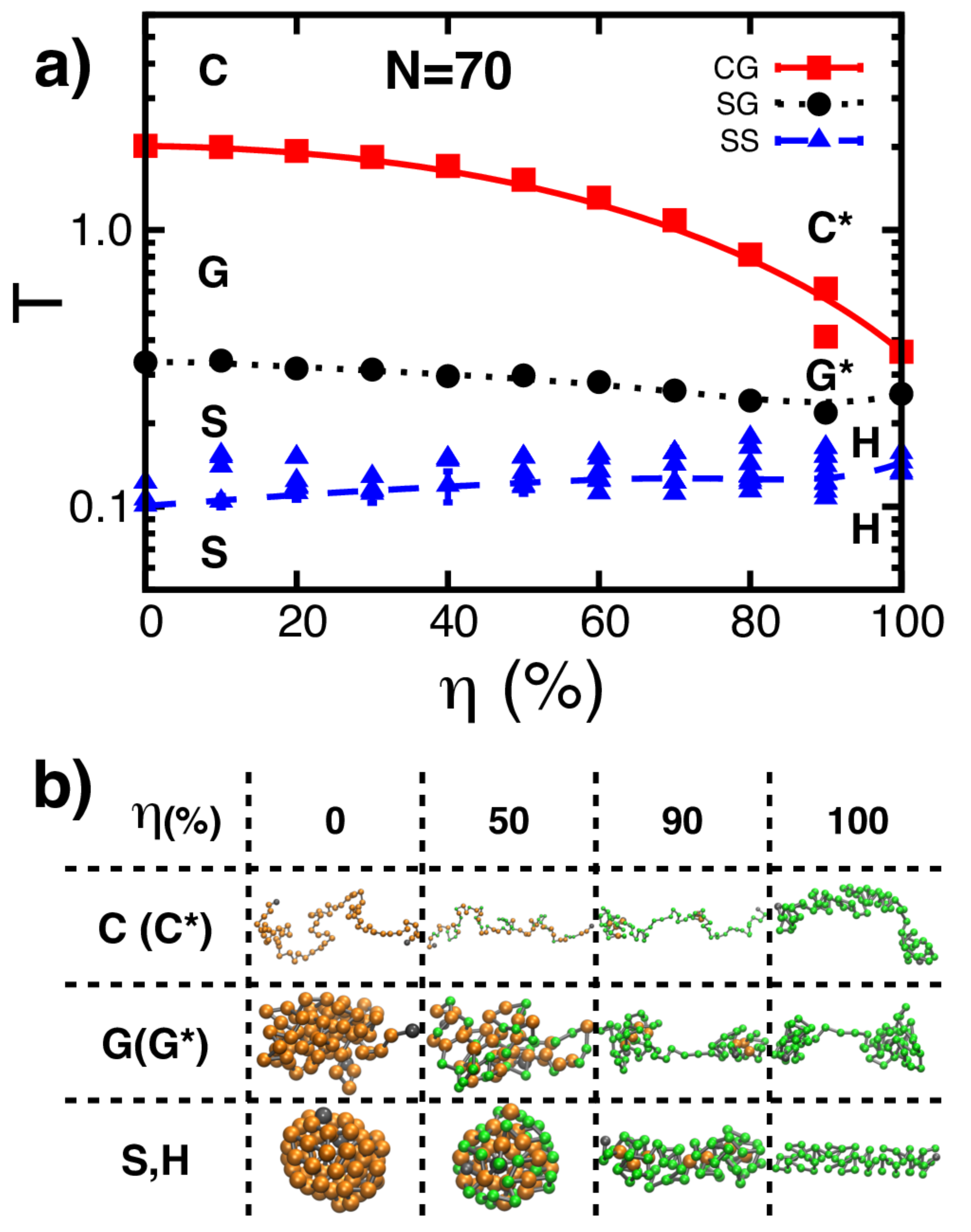}
    \caption{(Color online) $(a)$ Upper: Transition temperatures from peaks and shoulders in canonical fluctuations for chain length $N=70$, plotted versus charged monomer concentration $\eta$. Increasing $\eta$ lowers the temperature gap between the coil-globule (red squares) and the solid-globule (black circles) transitions. In contrast, the solid-solid (blue triangles) transition temperature has no apparent trend.     $(b)$ Bottom: Representative configurations for extended coil, $C,C^{\star}$, liquid globular, $G,G^{\star}$, solid globular, $S$, and solid helical, $H$, phases. Yellow spheres are for bare monomers and the green ones for monomers with screened Coulomb repulsion.}
    \label{Fig4_pd}
\end{figure}

The hyperphase diagram, depicted in Fig.~\ref{Fig4_pd}, is constructed from the peaks and shoulders in structural and thermal fluctuations in Fig.~\ref{Fig3_critical}. In Fig.~\ref{Fig4_pd} $(a)$ (upper panel), we show the transition temperature points with guiding lines for clarity (not to be interpreted as data interpolation) for $N=70$ monomers and varying $eta$, and in Fig.~\ref{Fig4_pd} $(b)$ (lower panel), we provide representative configurations for each phase. The coil-globule, \mbox{CG}, transition temperatures (red squares) are obtained from the maxima of the smooth peaks in $\langle \mathrm{d} R_{\mathrm{gyr}}/\mathrm{d}T\rangle$. The temperatures of the solid-globule, \mbox{SG}, transition (black circles) are obtained from the maxima of the sharp peaks in $\langle  C_{\mathrm{V}}/N\rangle$. The solid-solid, \mbox{SS}, transition (blue triangles) temperatures are collected from very low $T$ peaks in energetic and structural fluctuations. High thermal activity may be associated with folding/unfolding into a non-trivial topology (e.g., like knots, loops, and entangled chains)~\cite{Rafaello2015}. These interesting topological transitions can be explored with the inclusion of specific bond-exchange MC moves~\cite{Schnabel}, which were not included in our studies as they violate the quenched disordered statistics

Red squares are coil-globule (or pearl-necklace at high $\eta$), \mbox{CG}, black dots are for freezing solid-globule (or pearl-necklace to helical at high $\eta$), \mbox{SG}, and the solid-solid, \mbox{SS}, transition is represented by blue triangles. The lines are only guide to the eyes. We present equilibrium configurations for each structural phase in Fig.~\ref{Fig4_pd}-$(b)$.

Top figure shows the hyperphase diagram at the $T,\eta$ plane for $N=70$ monomers. Lines are guide to the eyes. Error bars are smaller than the symbols when not visible.  The region labeled $C,C^{\star}$ is a coil phase, although at higher $\eta$ the polymer starts forming agglomerates. The region $G$ corresponds to a globular phase. For $\eta>80\%$, a two-globule structure $G^{\star}$ is observed (See the figure in the bottom).  At very low T, the solid globule, $S$ phase, is the equilibrium configuration for $\eta <90\%$. Above this concentration, a zig-zag structure resembling a helix, $H$, appears. $(b$) Representative configurations for each structural phase mapped. Yellow spheres are for bare monomers and the green ones for monomers with screened Coulomb repulsion.

Four different regions are identified in the $(T,\eta)$ plane. The coil ($C,C^{\star}$) phase occurs at high temperature for all $\eta$ values. It is a random vapour-like  and structureless extended phase, dominated by entropic interactions. While at low $\eta$ the polymer is entropically driven and configurations $C$ are unstructured, at high charge concentration ($C^{\star}$ configurations), local agglomerates are formed. These ``blobs'' are similar to the pearl-necklace conformations of polyelectrolytes predicted both theoretically~\cite{Dobrynin1996} and experimentally~\cite{JACS2002}. By lowering the temperature, the polymer undergoes a transition at $T_{\text{CG}}$, which decreases when $\eta$ increases. The globular phase ($G,G^{\star}$) is characterized by the collapse of the polymer chain. When the charged monomers concentration increases, the polymer tends to trap the neutral monomers (yellow spheres) by surrounding them with the charged ones (green spheres), eventually preventing the chain to collapse into a single globule, clearly seen in Fig. \ref{Fig4_pd} $(b)$ for $\eta = 90\%$. The formation of two-droplets at $\eta=90\%$ is consistent with the two peaks observed at $\mathrm{d}\langle R^2_{\mathrm{g}}\rangle/\mathrm{d}T$ in Fig.~\ref{Fig3_critical}-$(g)$. Here we see an attempt to minimize both surface and enthalpic energies, with a bigger contribution from screened Coulomb interactions. As there exist a maximum stable net charge in a single blob (Rayleigh charge), when this maximum charge is reached and the two blobs approach, the effective Coulomb repulsion between newly formed blobs increases and the pearl-necklace unfolds. By lowering the temperature, the chain freezes into a stable zig-zag configuration.

For $\eta=100\%$ the two-droplets configuration at the liquid phase is not stable, and small random blobs are eventually folded and unfolded. This phase is characterized by partial collapses of chain segments, until the polymer freezes into a well-organized zig-zag structure at $T_{\mathrm{SG}}$. The solid phase, $(S,H)$, has two different representative structures. For $\eta<90\%$, the globular liquid structure freezes into a solid-globule formed by spherical layers. As for $N=70$, we are far from the ``magic numbers'' (the closest ones being $55$ and $147$), where full icosahedron layers are formed~\cite{seatonPRE}, and due to the quenched disorder, no crystalline structure is expected. For $\eta\ge 90\%$, the solid-globular freezing is prevented due to strong electrostatic repulsion. The more energetically favorable structure resembles a rod, but with twisted branches as found in proteins~\cite{Dill1991}. The twist is a natural way to minimize both the screened Coulomb repulsion and LJ attraction as predicted in Ref.~\cite{Science2005}. This structural transition has a high thermal activity, identified by the peak at $\langle C_V\rangle$  on Fig. \ref{Fig3_critical} $(h)$. Interestingly, the solid $S$ phase has intense thermal activity, as seen by the numerous blue triangles at $\eta=80,\,90$ and $100\%$, due to reorganizations of the solid helical phase. Transition temperatures for the bare polymer $T_{\text{CG}}\left(\eta=0\right)= 2.00(1)$ and $T_{\text{SG}}\left(\eta=0\right)= 0.33(1)$ are in excellent agreement with those predicted by scaling-laws from Ref.~\cite{seatonPRE} where they found $T_{\text{CG}}=1.977(4)$ and $T_{\text{SG}}=0.33(1)$ for $N=70$. The dashed (blue) line shows the solid-solid, \mbox{SS}, transition. We find $T_{\text{SS}}\left(\eta=0\%\right)=0.1008(1)$, follows a straight line in the entire range of charge concentration. For a fully charged polymer, $T_{\text{SS}}\left(\eta=100\%\right) = 0.144(8)$.

\subsection{Classification of structural transitions}
\label{Struct_Transitions}
\noindent

The classification of structural transitions in finite systems is not a simple task, as traditional techniques require that the ratio between surface and volume entropy contributions tend to zero. Therefore, more sophisticated methods as for instance, Fisher zeros~\cite{fisher1998renormalization}, Energy Probability Distribution zeros~\cite{costa2017energy} or microcanonic analysis are meant to overcome this limitation. As both methods deal with the derivative of the microcanonical entropy, instead of the free energy, they are less sensitive to the finiteness of the system. The microcanonical properties are derived from the probability distribution $P\left(E,T\right)$ at the transition temperature~\cite{BinderIJMPC}. First-order transitions reveal a bimodal probability distribution, where each peak represents the energy of maximum probability for each phase above and below the transition, let as say $B$ and $A$, respectively. The position of each peak gives the equilibrium energy of the respective phase (for a fixed temperature $T$), $E_A$ for the phase $A$ and $E_B$ for the phase $B$. In a first order transition, the system has an equal probability to be found in each phase, then the phases coexists. The distance in energy between the peaks, $\Delta Q = \left|E_B - E_A\right|$, is the latent heat. The bimodal form is directly connected with the microcanical analysis for a first order transition~\cite{MichaelBook},as discussed in the following. \\

\begin{figure}[htb!]
\centering
    \begin{minipage}[b]{0.45\linewidth}
    \includegraphics[width=\linewidth]{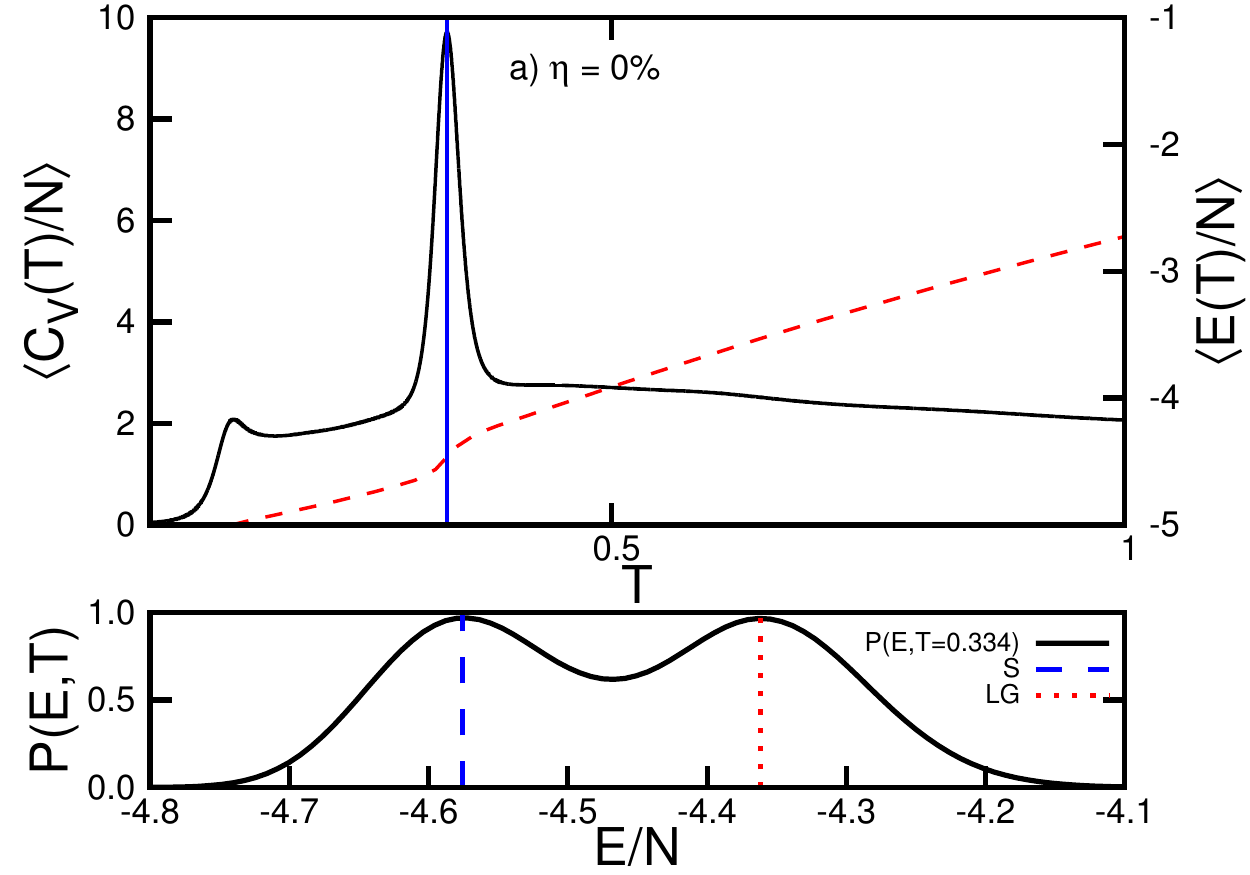}
\end{minipage}
\begin{minipage}[b]{0.45\linewidth}
    \includegraphics[width=\linewidth]{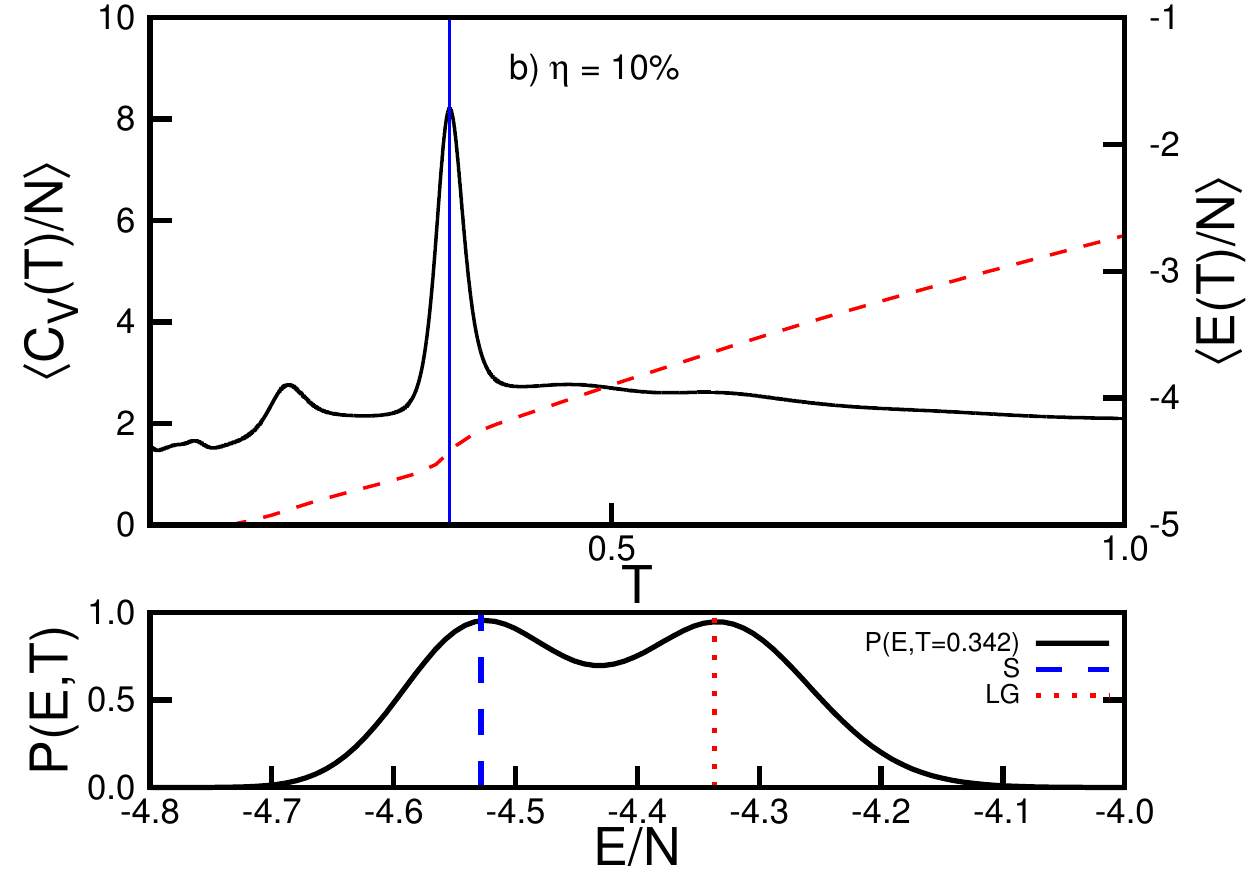}
\end{minipage}
\begin{minipage}[b]{0.45\linewidth}
    \includegraphics[width=\linewidth]{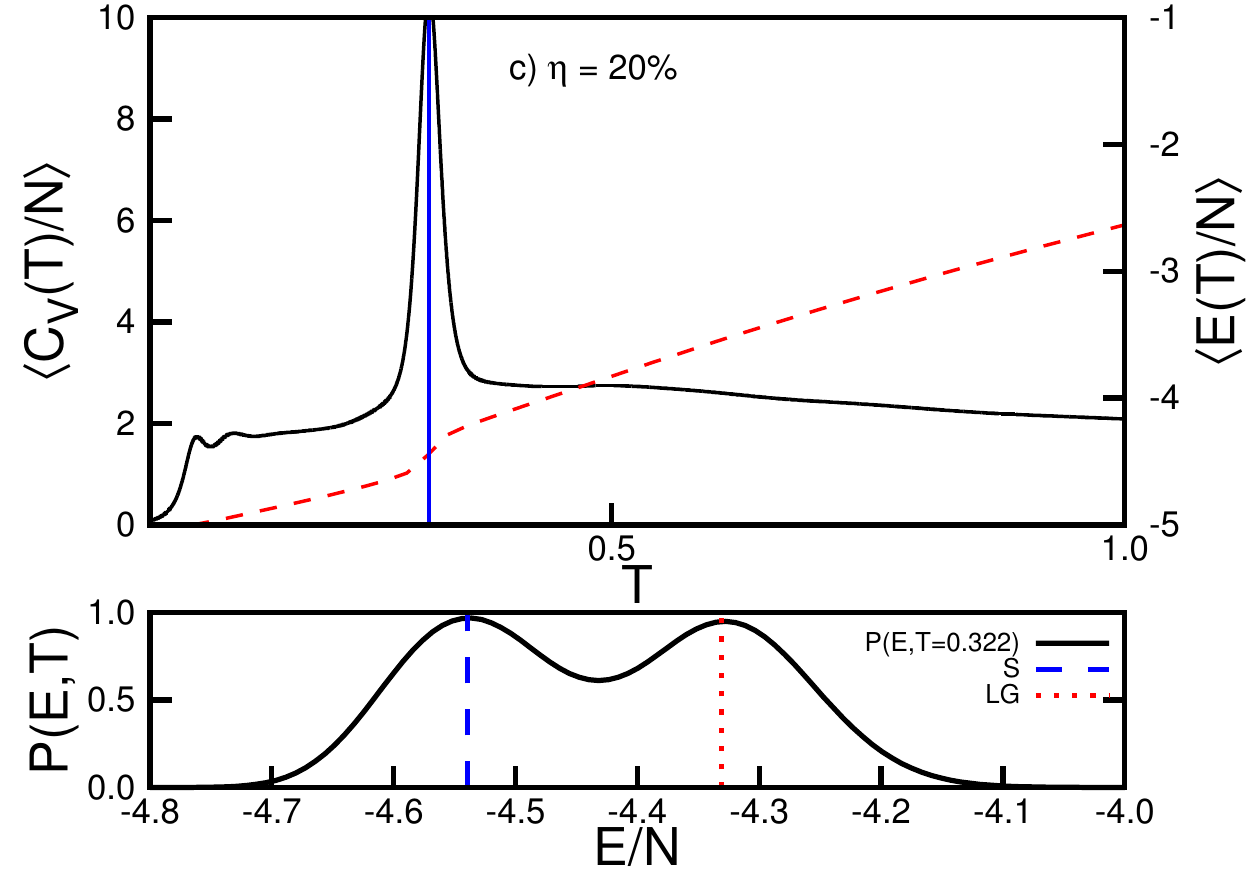}
\end{minipage}
\begin{minipage}[b]{0.45\linewidth}
    \includegraphics[width=\linewidth]{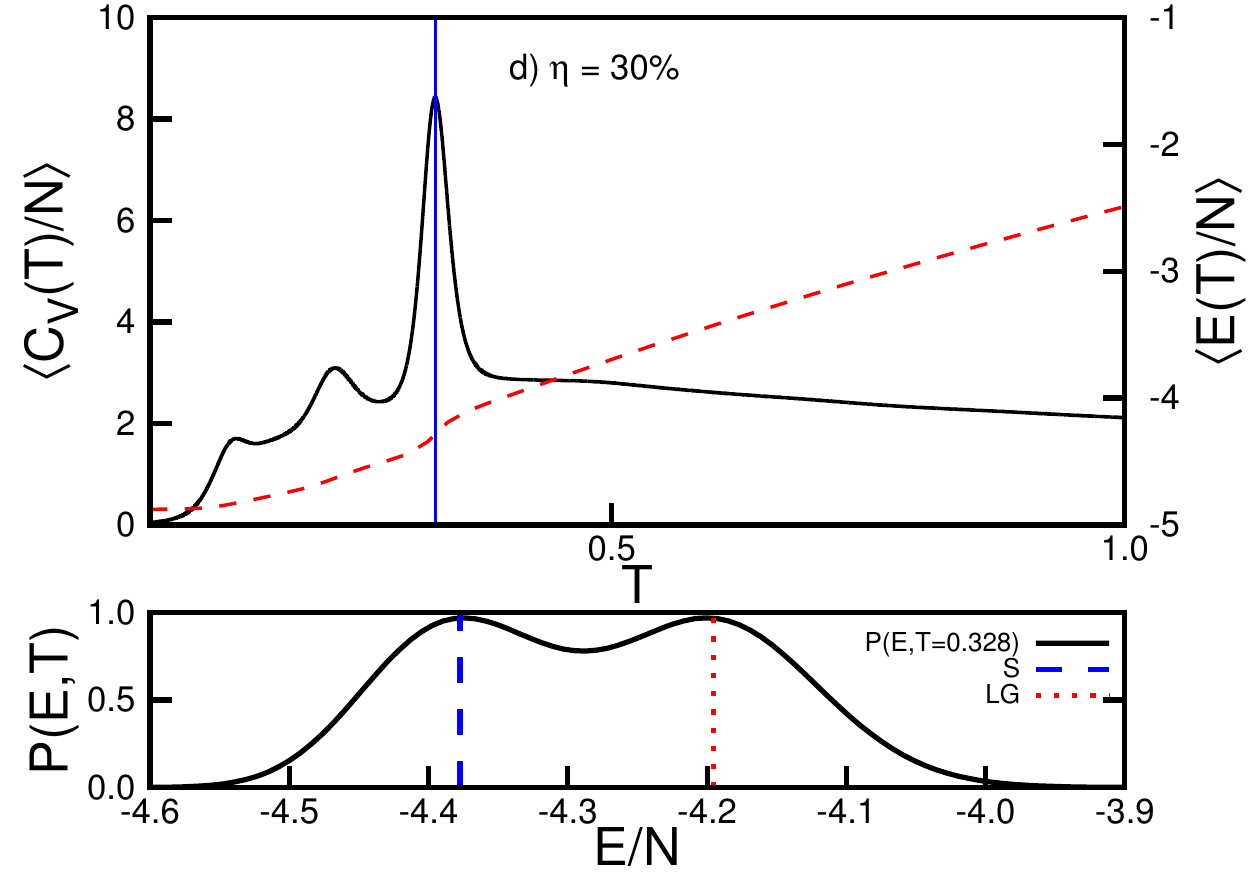}
\end{minipage}
\begin{minipage}[b]{0.45\linewidth}
    \includegraphics[width=\linewidth]{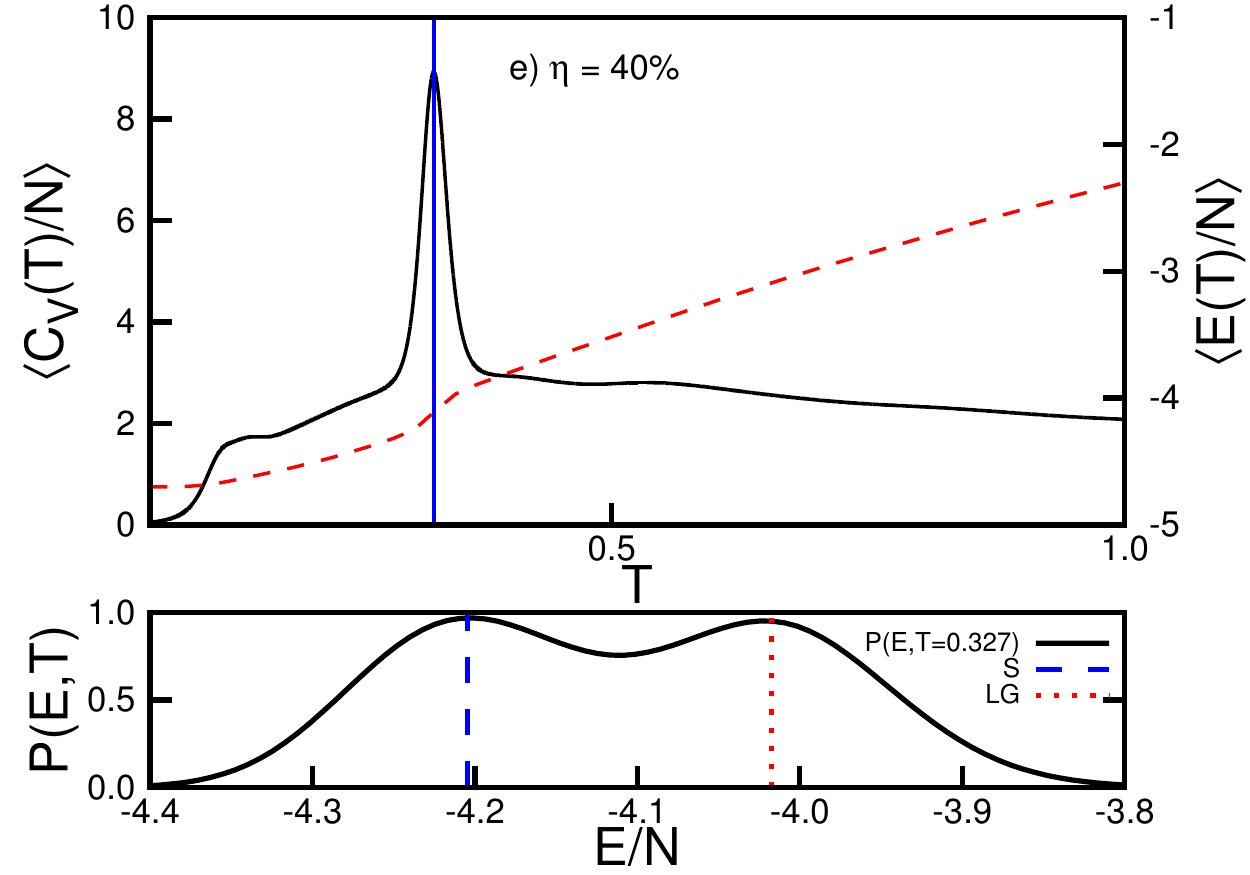}
\end{minipage}
\begin{minipage}[b]{0.45\linewidth}
    \includegraphics[width=\linewidth]{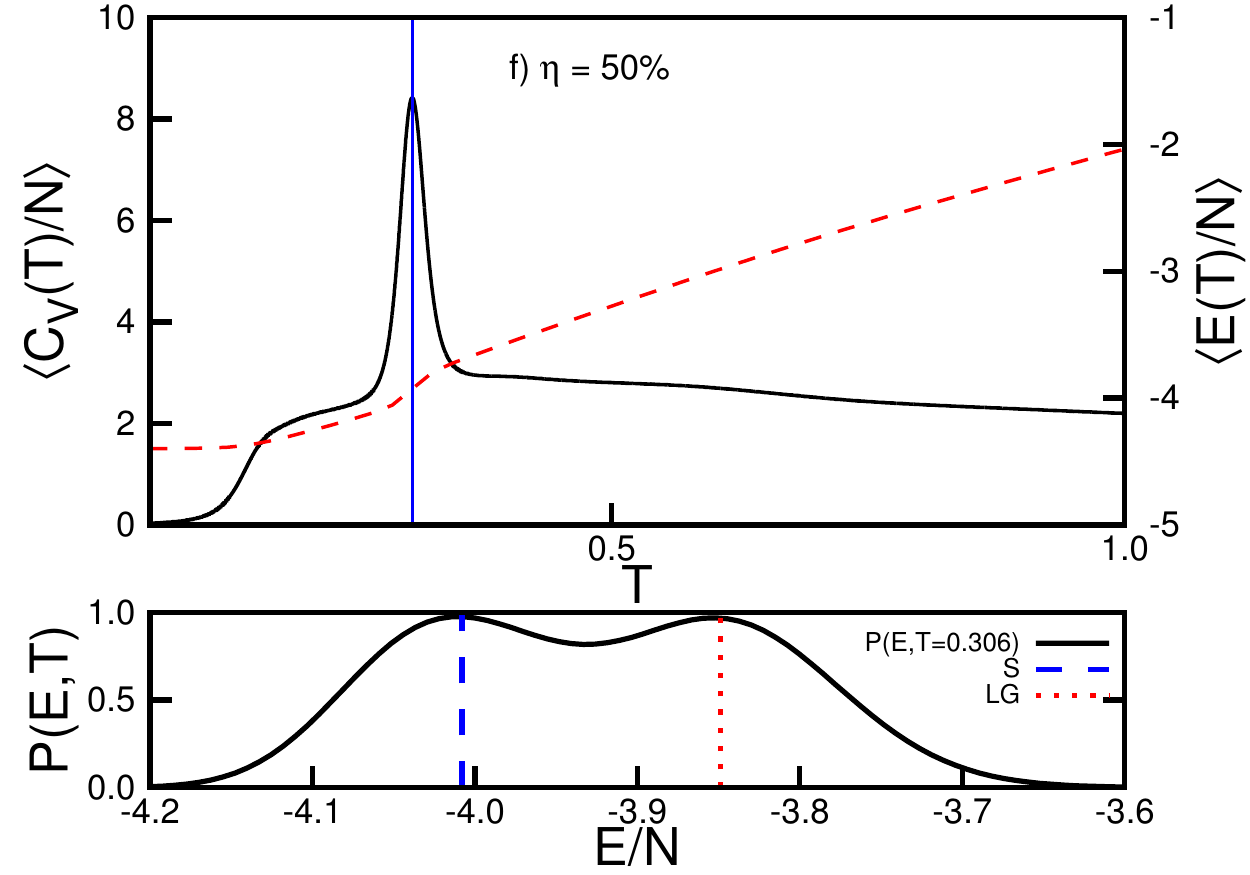}
\end{minipage}
\begin{minipage}[b]{0.45\linewidth}
    \includegraphics[width=\linewidth]{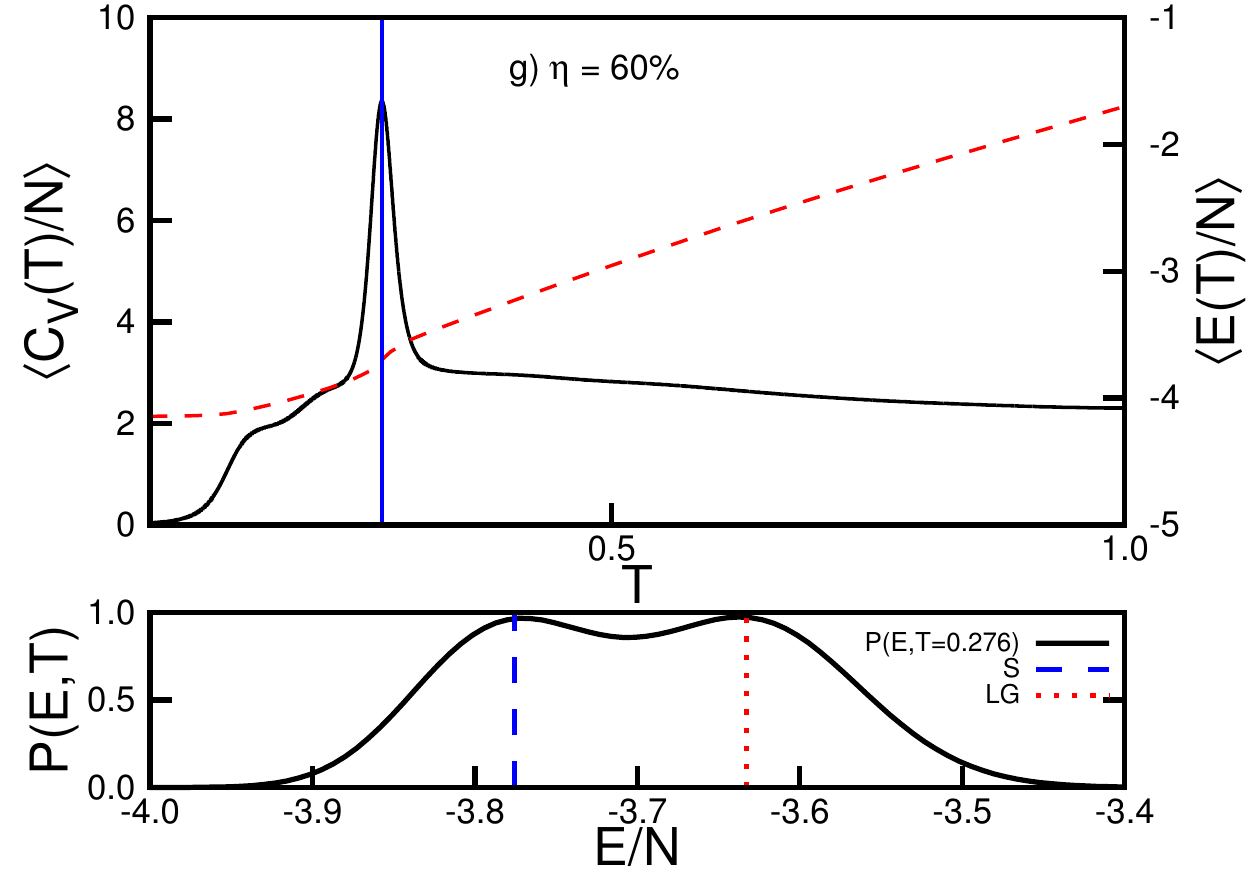}
\end{minipage}
\begin{minipage}[b]{0.45\linewidth}
    \includegraphics[width=\linewidth]{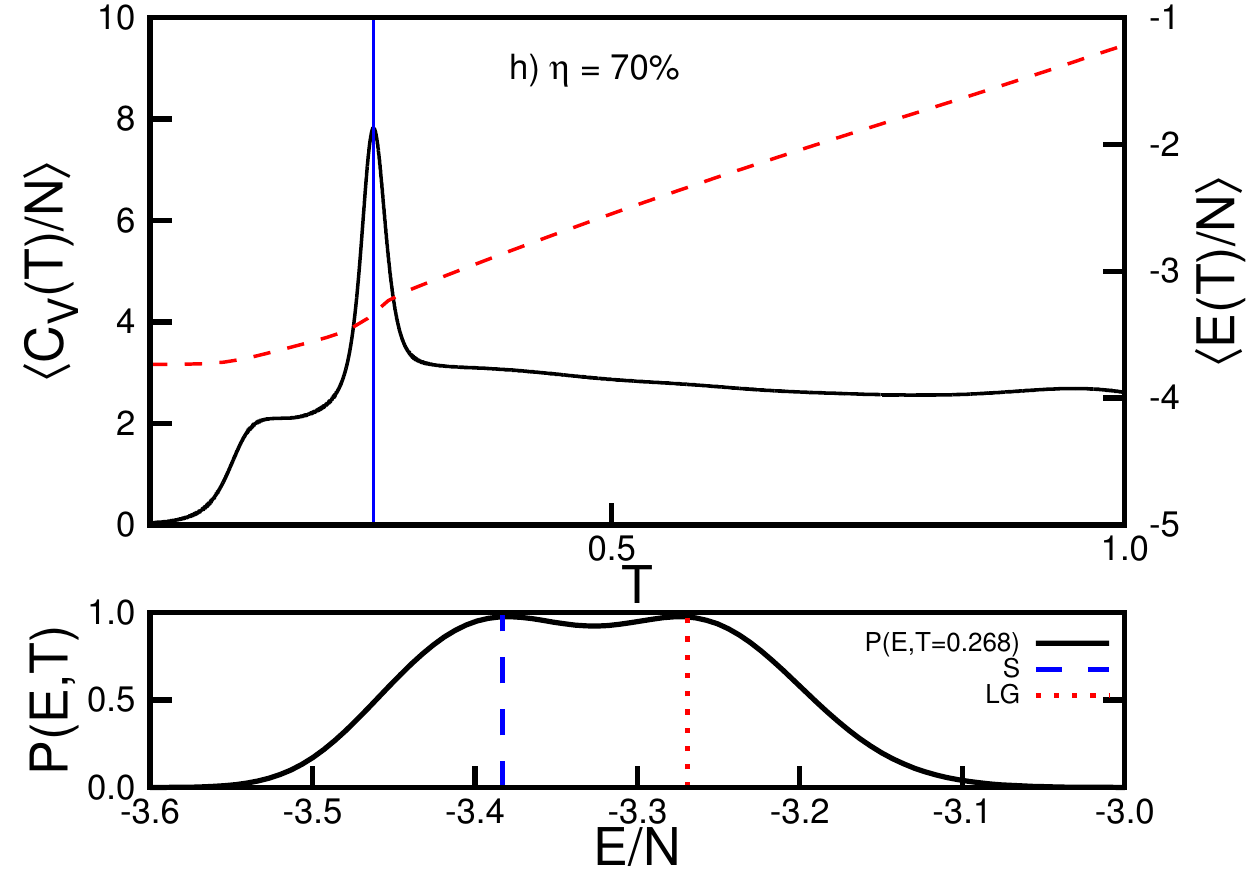}
\end{minipage}
\begin{minipage}[b]{0.45\linewidth}
    \includegraphics[width=\linewidth]{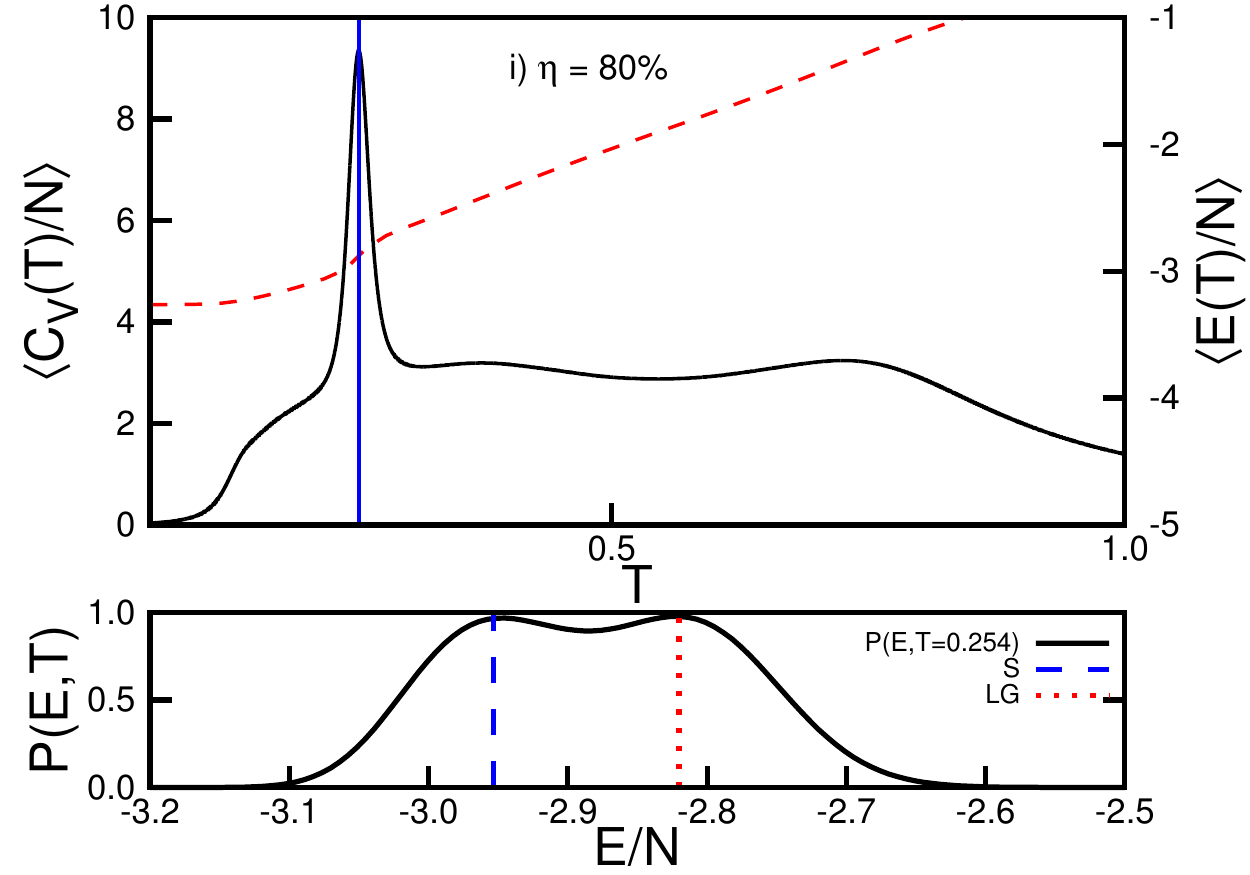}
\end{minipage}
\begin{minipage}[b]{0.45\linewidth}
    \includegraphics[width=\linewidth]{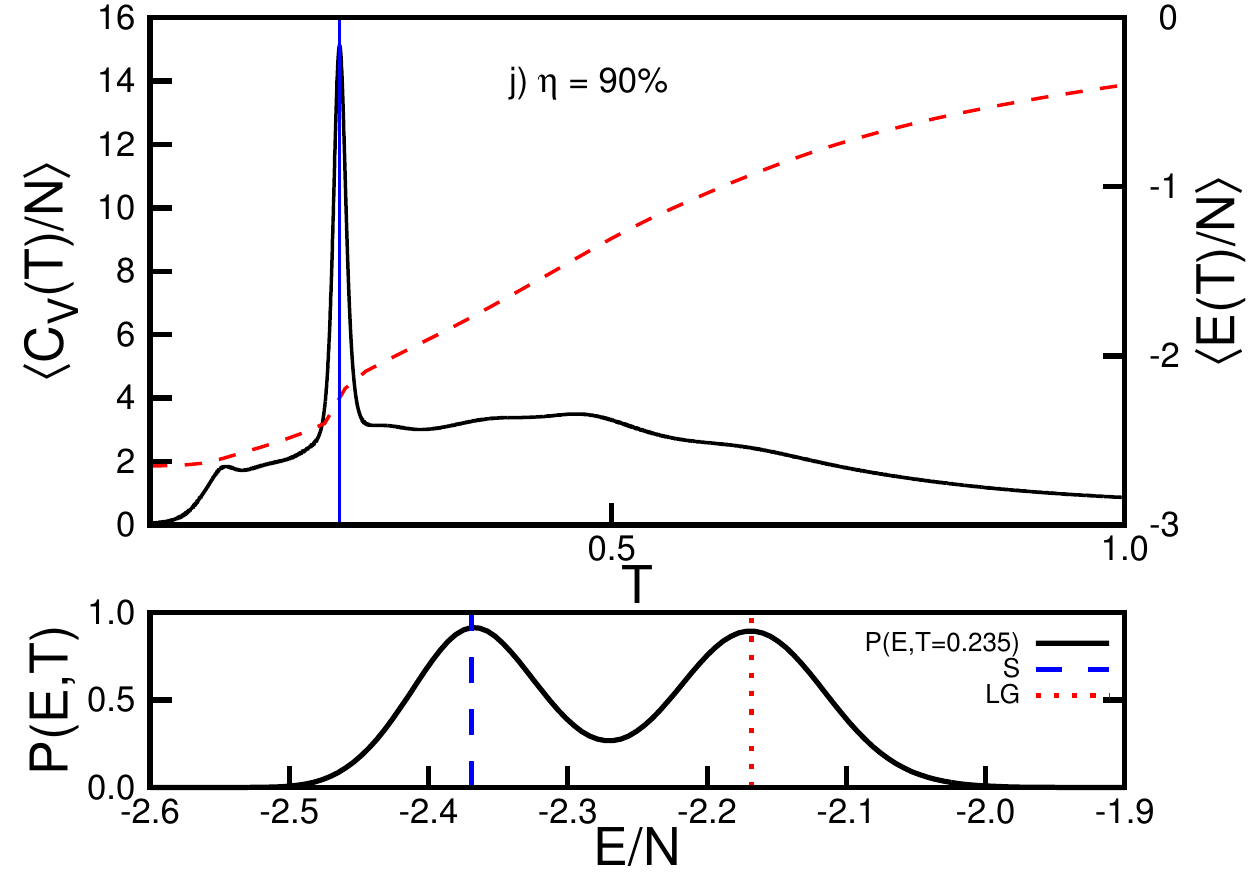}
\end{minipage}
    \caption{\label{fig:transicao0}(Color online) Specific heat (black) and energy (red) as a function of temperature for $\eta = 0, 10, \ldots 80$ and $90\%$. Vertical blue line marks the transition temperature. Probability $P(E,T)$ is shown below each plot, where the first (blue) and second (red) peaks are for the equilibria  temperature for solid and liquid phases, respectively.}
\end{figure}

In Fig. \ref{fig:transicao0} we show the analysis of the structural transitions between solid and globular phases  in the range $\eta=0,...,90\%$ (panels $a-j$). Those are consistent with first-order classification. The transition temperature is identified as a vertical blue line at the peak position. Also, we show the canonical energy, $\langle E\left(T\right)\rangle$, as a red dashed line (the scale is read at the right vertical axis). The change in the energy curvature, $\langle E\left(T\right)\rangle = -\frac{\partial \log Z}{\partial \beta}$, identifies the transition canonical temperature.

In the lower panels we present the probability, $P\left(E,T\right)$  at $T=T^{tr}$, with a bimodal distribution. The left peak (lower energy) marks the equilibrium energy of the solid phase (dashed vertical blue line), while the right peak locates the respective globular phase (vertical dotted red line). The depth of the valley between both phases is proportional to the free energy barrier required for the transition to happen ~\cite{BinderIJMPC}. Within this framework, we can address some questions left unanswered in the canonical analysis. First, an increase in $\eta$ results in a decrease in the depth of the valley, up to $\eta=80\%$. For $\eta=90\%$, the depth of the valley is even more pronounced than the one at zero concentration. However, we can see that the transition happens between a solid toroidal and a liquid globular phase (Fig. \ref{Fig4_pd}). This transition is, therefore, of first order, as we can see at the lower panel in Fig. \ref{fig:transicao0} $j)$. This result shows how a zig-zag very stable configuration naturally emerges from simple two-body potentials, in comparison with previous models where thee-body (bending angles) and four-body (dihedral angles) potentials are required~\cite{MichaelPRL}.

\begin{figure}[htb!]
\centering
\begin{minipage}[b]{0.45\linewidth}
    \includegraphics[width=\linewidth]{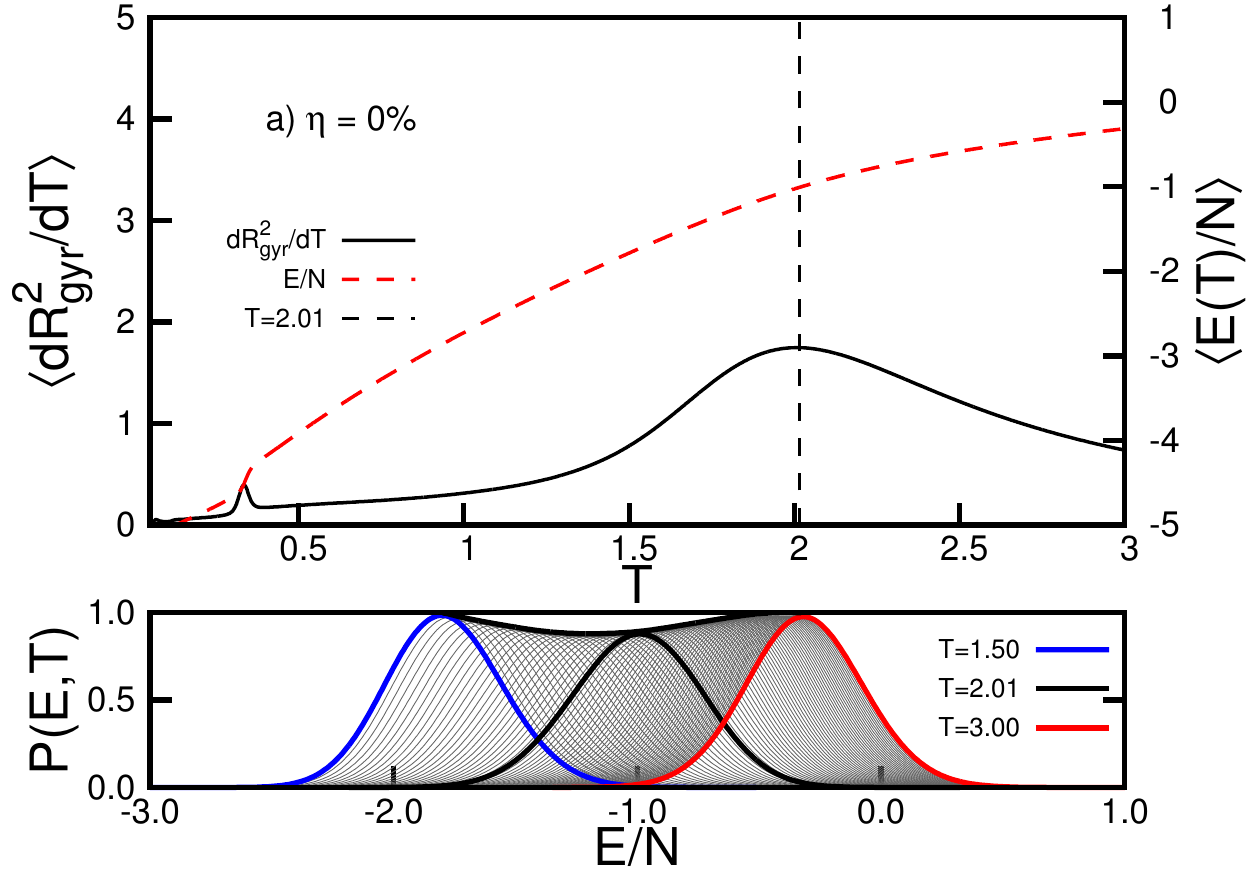}
\end{minipage}
\begin{minipage}[b]{0.45\linewidth}
    \includegraphics[width=\linewidth]{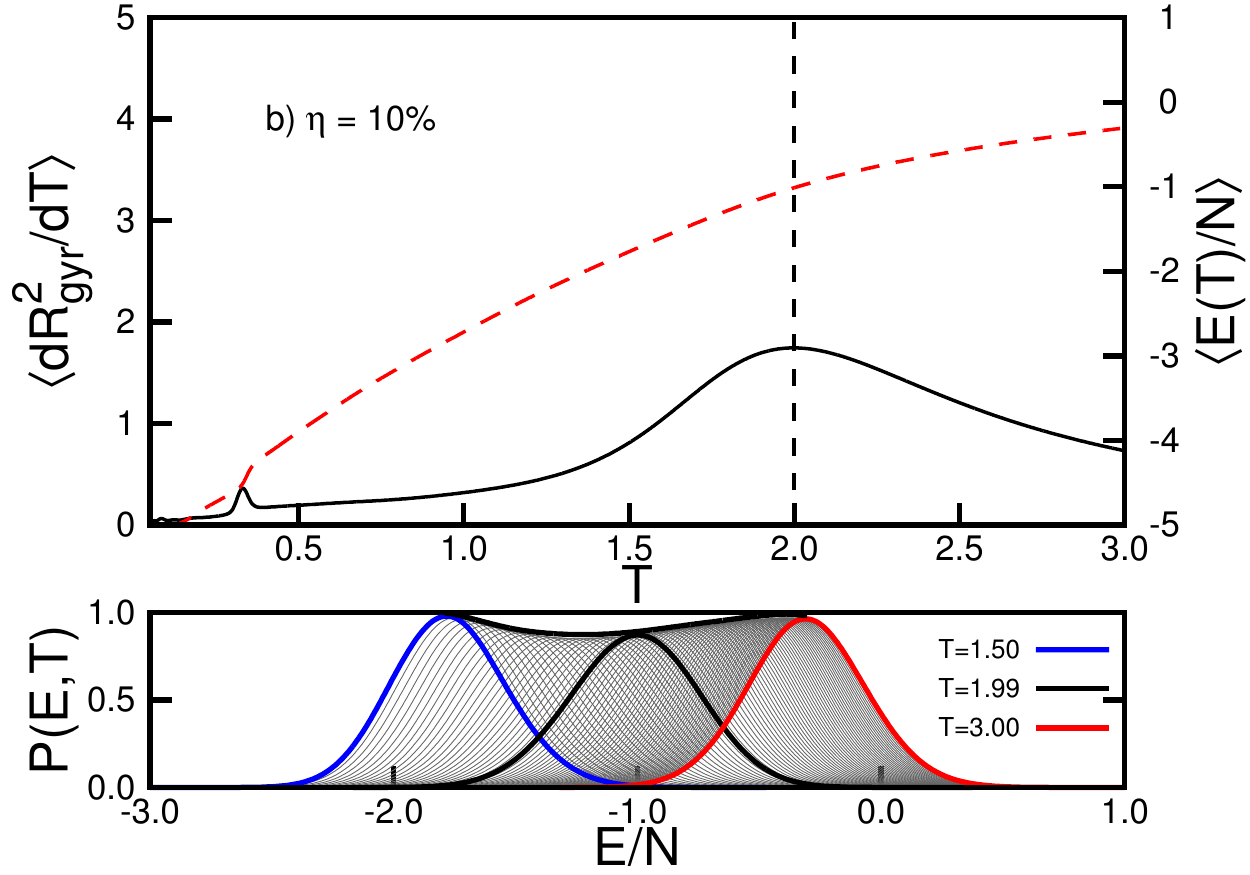}
\end{minipage}
\begin{minipage}[b]{0.45\linewidth}
    \includegraphics[width=\linewidth]{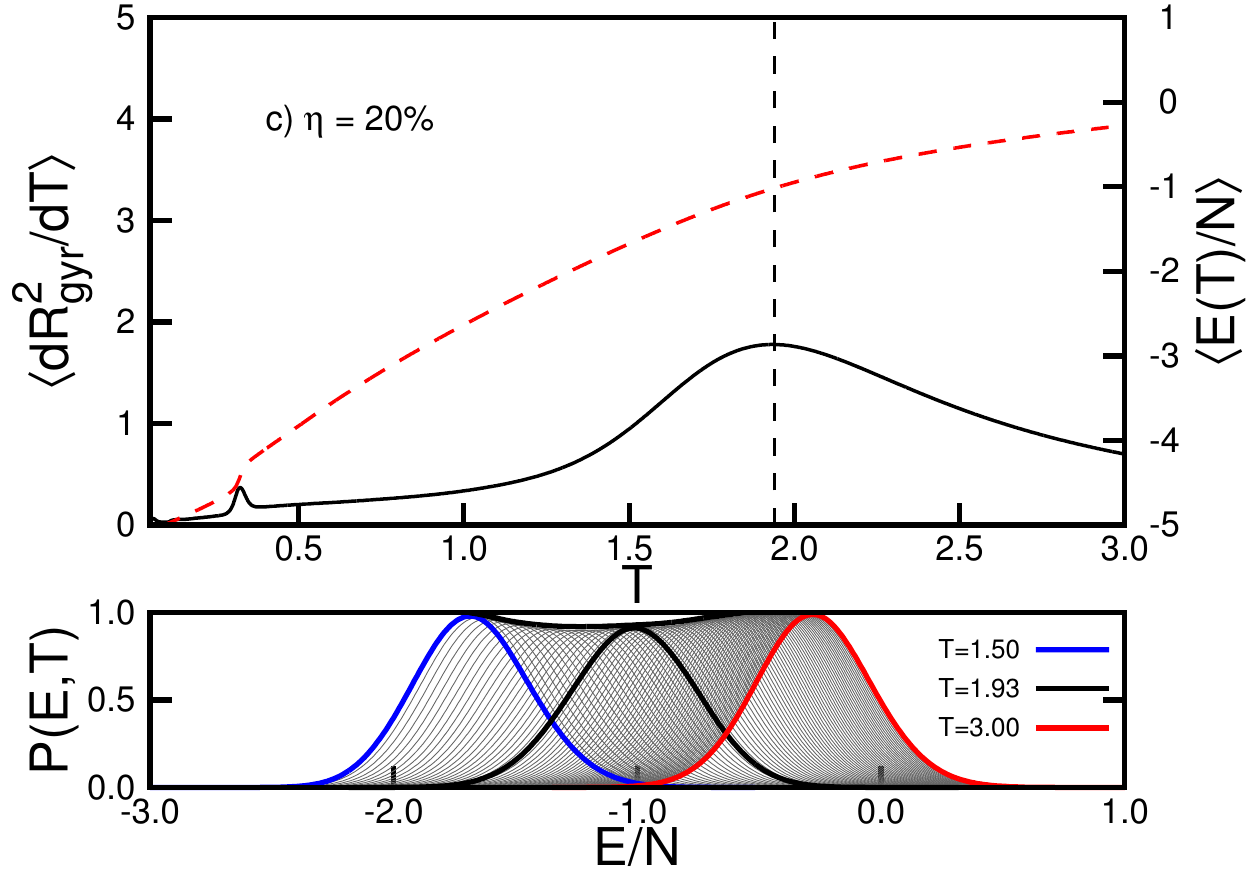}
\end{minipage}
\begin{minipage}[b]{0.45\linewidth}
    \includegraphics[width=\linewidth]{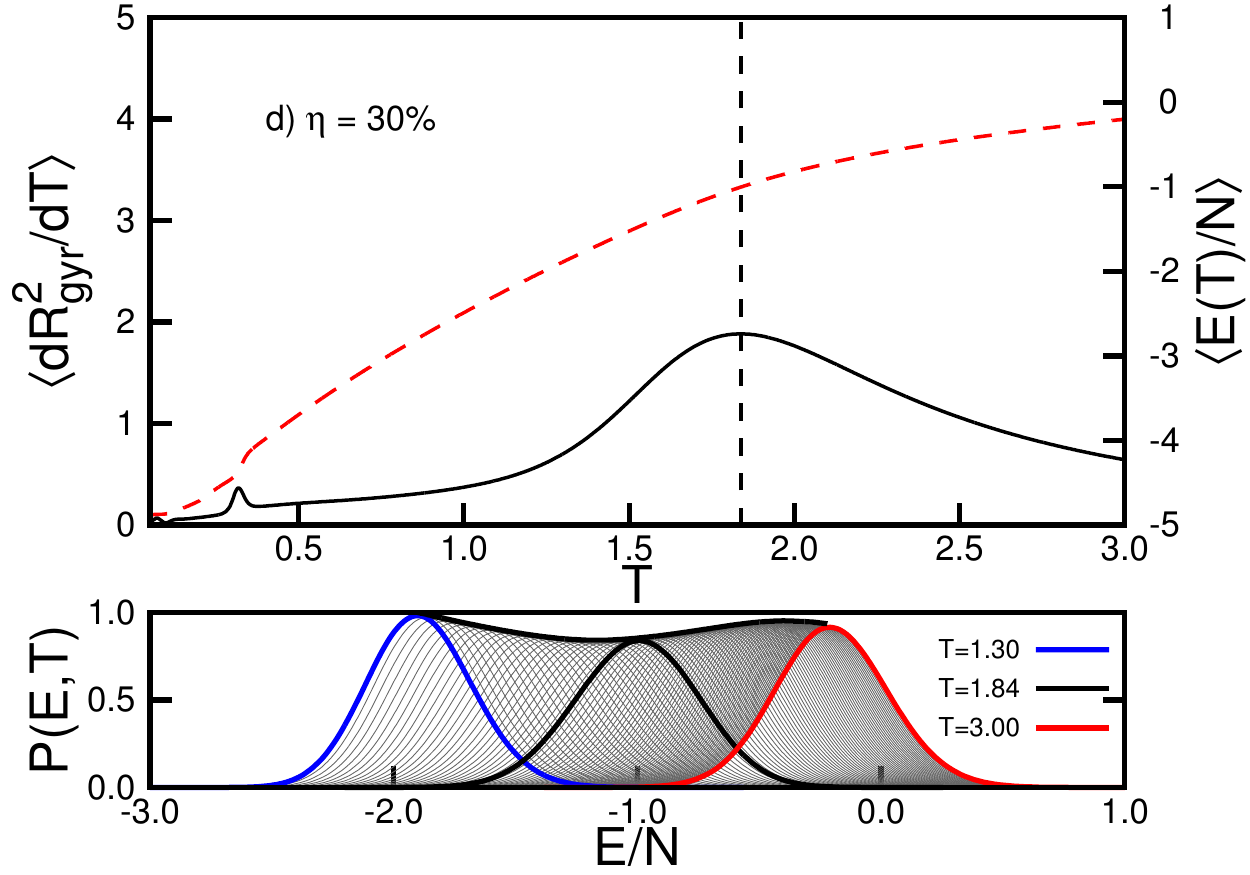}
\end{minipage}
\begin{minipage}[b]{0.45\linewidth}
    \includegraphics[width=\linewidth]{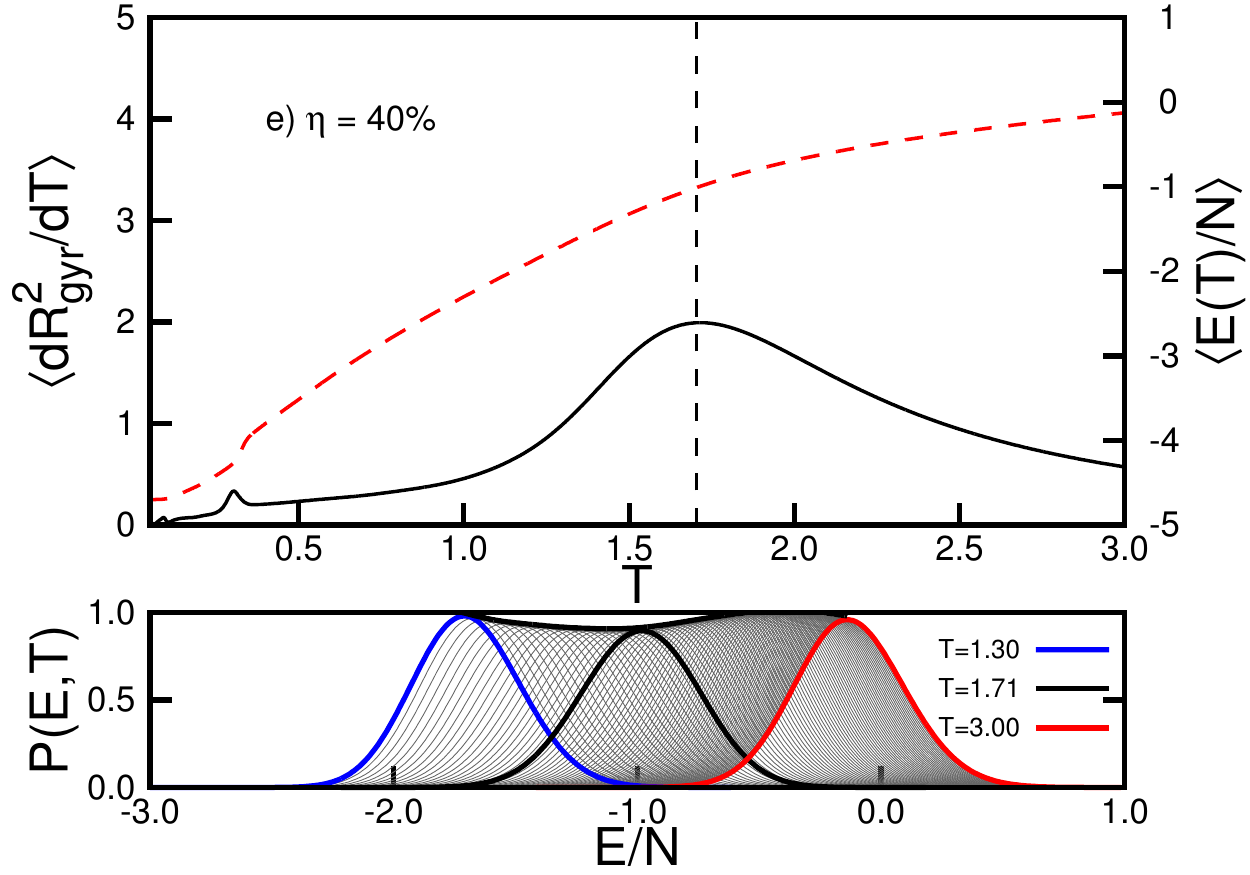}
\end{minipage}
\begin{minipage}[b]{0.45\linewidth}
    \includegraphics[width=\linewidth]{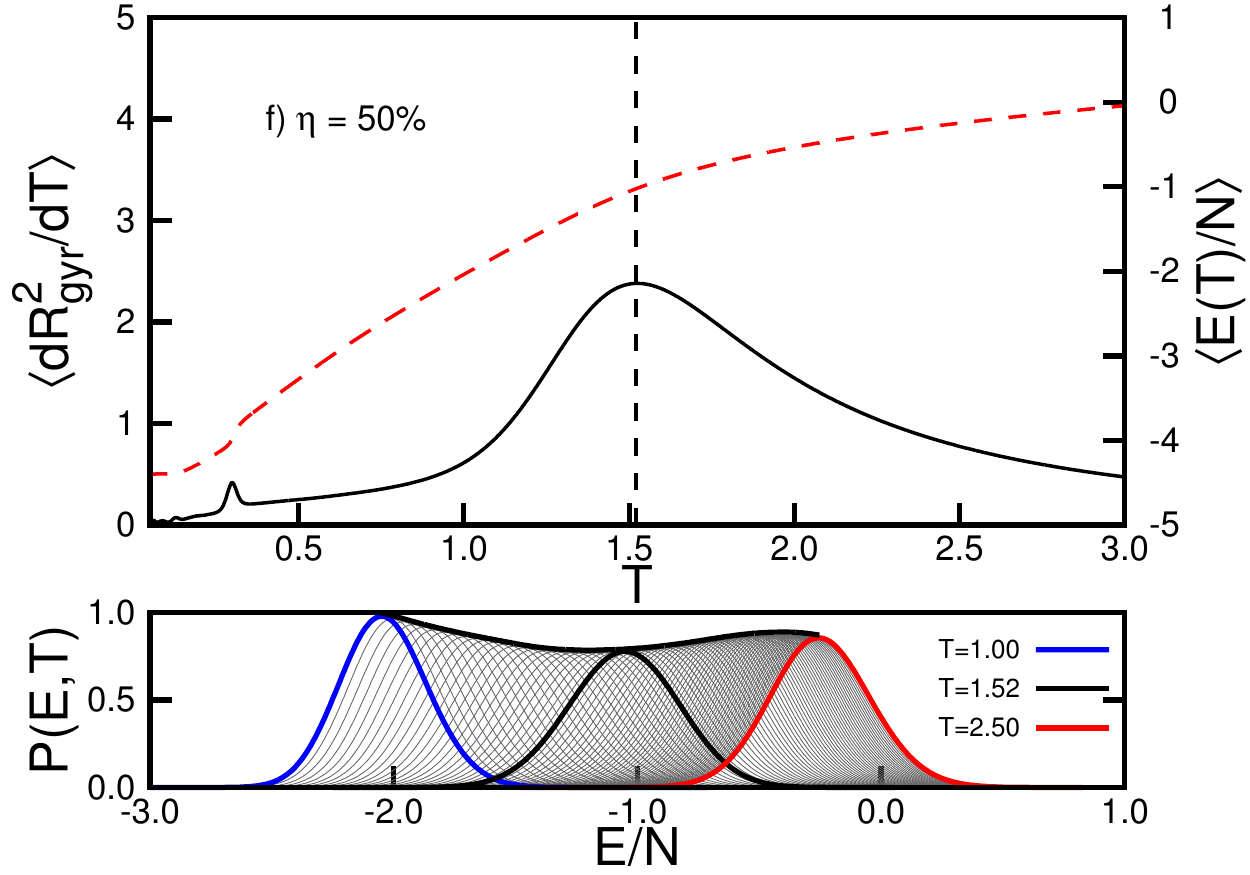}
\end{minipage}
\begin{minipage}[b]{0.45\linewidth}
    \includegraphics[width=\linewidth]{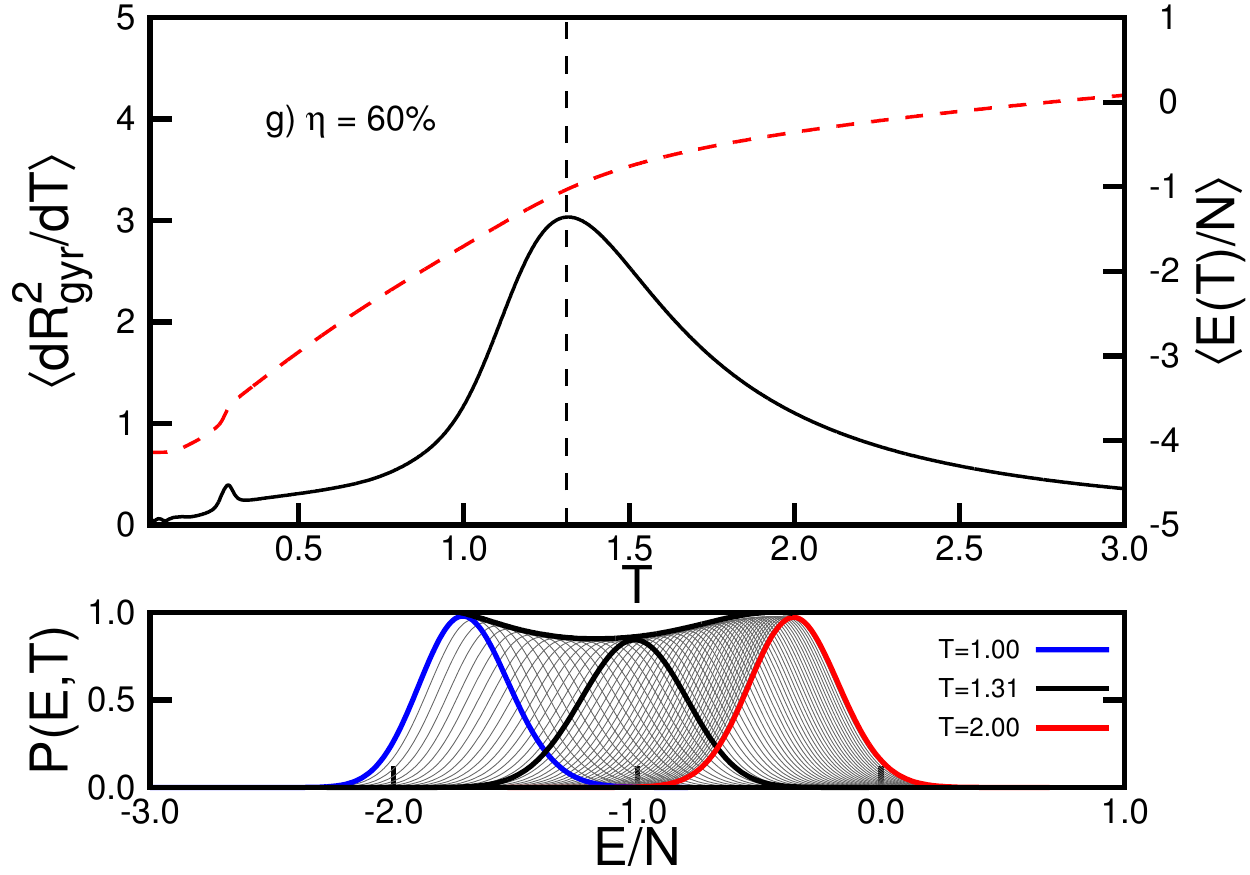}
\end{minipage}
\begin{minipage}[b]{0.45\linewidth}
    \includegraphics[width=\linewidth]{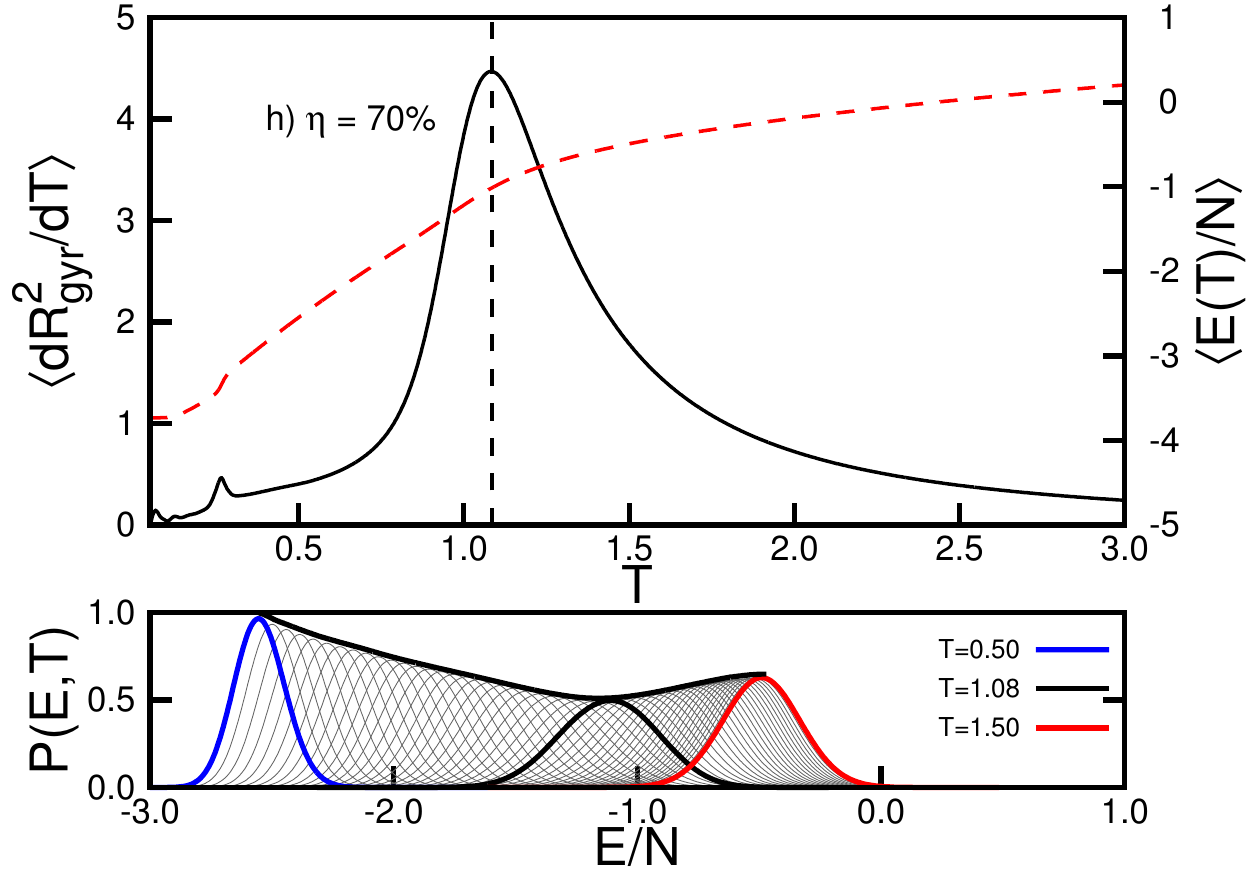}
\end{minipage}
    \caption{\label{fig:transicao2a0}(Color online) Fluctuations in the radius of gyration, $\langle\mathrm{d}R_{\mathrm{gyr}}^2\left(T\right)/\mathrm{d}T\rangle$, with a vertical dashed black line marking the temperature of the collapse transition. The probability, $P\left(E,T=T\right)$, in the vicinity of $T=T^{tr}$ is shown below each panel. The highlighted curves are the distributions for liquid globular (solid blue) and extended coil (solid red) phases, while the black solid line is positioned at the transition.}
\end{figure}

\begin{figure}[htb!]
\centering
\begin{minipage}[b]{0.49\linewidth}
\includegraphics[width=\linewidth]{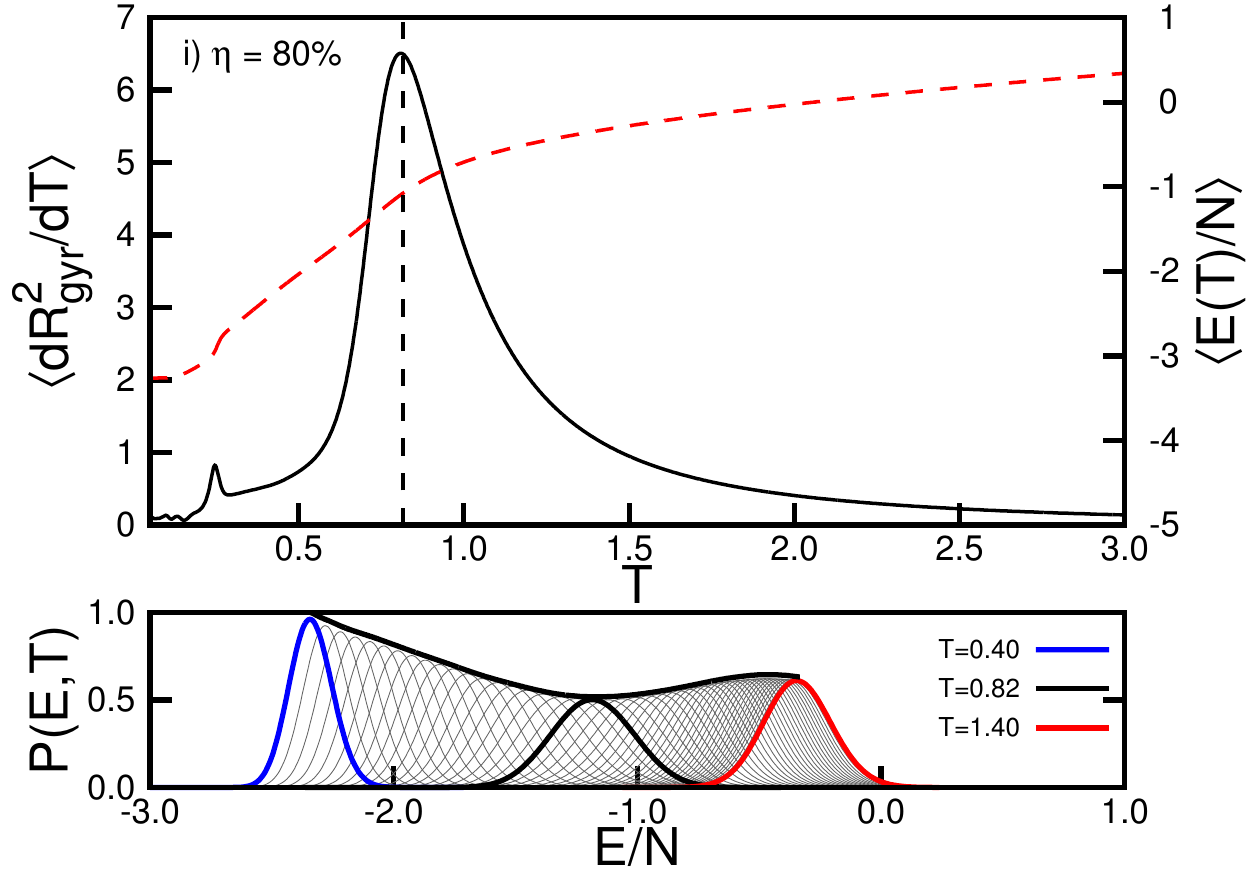}
\end{minipage}
\begin{minipage}[b]{0.49\linewidth}
\includegraphics[width=\linewidth]{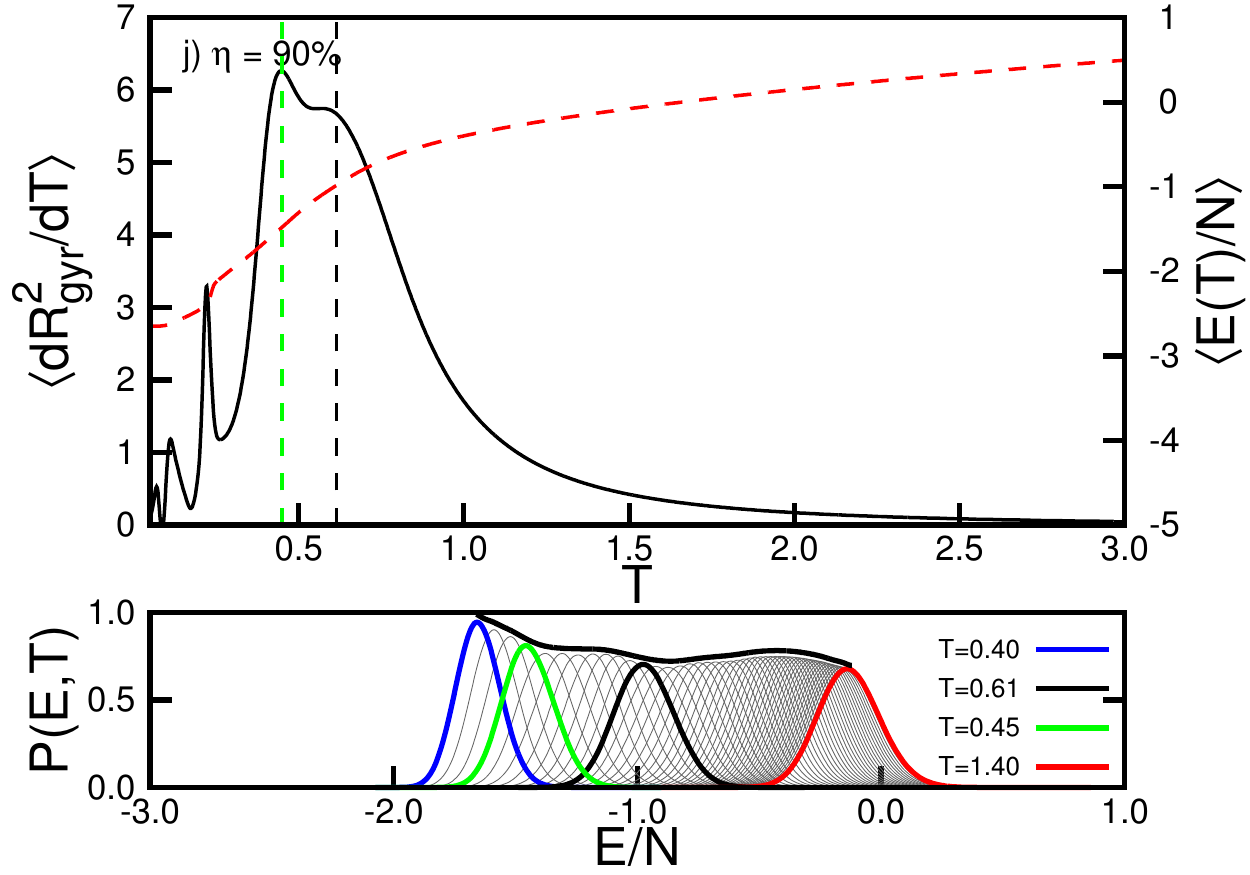}
\end{minipage}
\caption{\label{fig:transicao2a90}(Color online) The same as in Fig. \ref{fig:transicao2a0} for $\eta=80\%$ and $90\%$. the $\eta=80\%$ curve is essentially the same as for lower $\eta$, however, for $\eta = 90\%$ the collapse transition splits up in two, represented by the green and black curves.}
\end{figure}
\begin{figure}[htb!]
\centering
    \includegraphics[width=\linewidth]{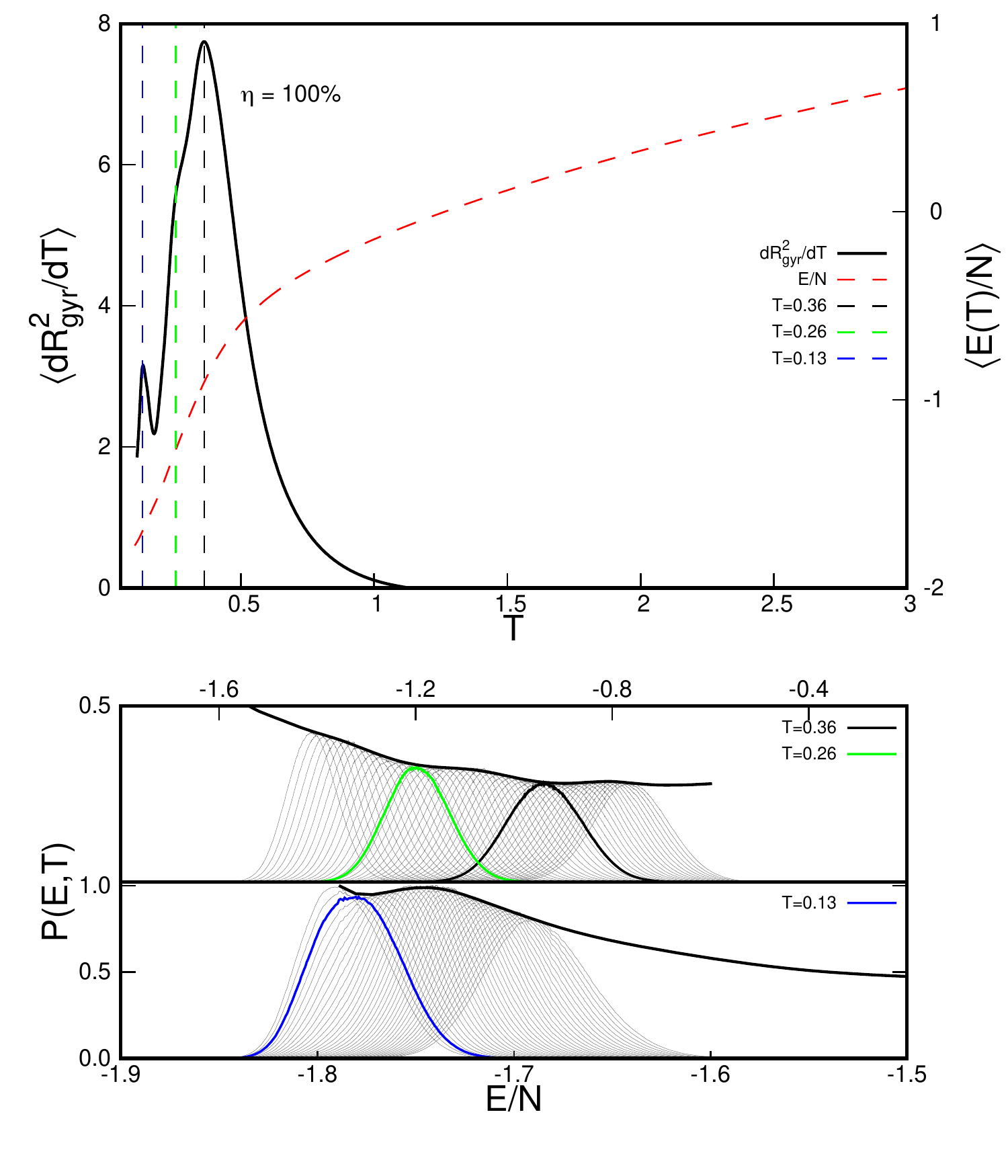}
\caption{(Color online) The same as in Fig. \ref{fig:transicao2a90} for $\eta=100\%$. We highlight two minima in the first plot and a third minimum at a lower temperature at $E/N \approx -1.8$.}
\label{fig:transicao2a100}
\end{figure}

In Figs. \ref{fig:transicao2a0} and \ref{fig:transicao2a90} we present an analysis of the continuous structural transitions already identified by the structural fluctuations (black solid line) and reinforced by the discontinuity in the canonical energy (red dashed line). Below those quantities we introduce the corresponding probability distribution $P(E,T)$ for a range of temperatures in the vicinity of the transition. The highlighted curves are for $T$ slightly below (blue curve) and above (red curve) $T_c$ (black curve). This representation allows one to identify a wrapping curve with a well defined minimum at $T_c$.  At high concentration ($\eta > 60\%$), the probability distribution becomes asymmetric, due to the energy difference between the extended and collapsed phases posed by the screening repulsive interaction. Nevertheless, the transition remains second-order. The first order transition is identified as a change in the concavity of the energy and a small peak in $\langle \mathrm{d}R_{\mathrm{gyr}}^2\left(T\right)/\mathrm{d}T\rangle$. At $\eta = 100\%$ the first order peak disappears, and three second order-like transitions show up as seen in Fig. \ref{fig:transicao2a100}. Recent works have shown structural transitions for single chain polyampholytes in agreement with our findings~\cite{EveraersPRL,WinklerPRE}. 

\section{\label{Conclusions}Conclusions}

In summary, we observed that polymer thermodynamics is directly affected by the introduction of monomers with screened Coulomb repulsion, modeled as a pairwise quenched disorder~\cite{Binder2011glassy}. The hyperphase diagram for $N=70$ monomers is reported in Fig.~\ref{Fig4_pd} $(a)$, showing a monotonic lowering of $T_{\mathrm{CG}}$ when $\eta$ increases. Our model matches previous results of $T_{\mathrm{CG}}$ for $\eta=0\%$~\cite{seatonPRE}. The system has three identified phases. A disordered extended phase, $(C,C^{\star})$, at $T>T_{\mathrm{CG}}$, a collapsed phase with globular and pearl-necklace configurations, $(G,G^{\star})$, at $T_{\mathrm{SG}}<T<T_{\mathrm{CG}}$ and a solid phase at $T<T_{\mathrm{SG}}$, with both solid globular, $(S)$, and helical, $(H)$, structures. Those phases are shown in Fig.~\ref{Fig4_pd} $(b)$.  At high charged monomers concentration, excluded-volume (EV) interactions become strong enough to prevent the collapse into a globule. At the same time, the transition remains entropically driven, as the non-bonded interactions do not directly restrict the degrees of freedom of the chain.~\cite{zierenberg2016dilute}. Interestingly, the polymer is stable in a helical configuration at the solid phase. The stability of the helix is a signature of a first-order transition, with a free-energy barrier most likely higher than the fluctuations of the surrounding environment. Indeed, the free-energy barrier sets the denaturation limit for the system. The calculation of free energies differences, especially in nucleation processes, has been a difficult task, due to the complexity and ambiguities of choosing a reaction coordinate at the conformational phase space. Sampling over the transition pathway requires accounting for the surrounding fluctuations that contribute to the probability distribution.

\begin{acknowledgments}
The authors gratefully acknowledge financial support from CNPq and FAPEMIG (Brazilian Agencies) under Grants CNPq 402091/2012-4 and FAPEMIG RED-00458-16.

\end{acknowledgments}

\bibliography{paper_diagram}

\end{document}